\documentclass[pra,showpacs,groupedaddress,notitlepage,
superscriptaddress,floatfix,nofootinbib,twocolumn]{revtex4-2}

\usepackage[english]{babel}
\usepackage{amsmath}
\usepackage{amssymb}
\usepackage{amsthm}
\usepackage{amsfonts}
\usepackage{braket}%Dirac Notation in QM
\usepackage{cases}
\usepackage{color}
\usepackage{xcolor}
\usepackage[font=small,labelfont=bf,justification=raggedright]{caption}
\usepackage{graphicx,subcaption}% hy77 Include figure files
\usepackage{multirow}
\usepackage{float}
\usepackage{mathtools}
\usepackage{ragged2e}
\usepackage{bm}% bold math
\usepackage{hyperref}
\hypersetup{
    colorlinks=true, 
    linkcolor=cyan,
    citecolor=magenta, 
    filecolor=magenta, 
    urlcolor=cyan,
    runcolor=cyan
}
\usepackage[capitalise]{cleveref}
\usepackage{booktabs} 
\usepackage{enumitem}
\usepackage{mathrsfs}

\newcommand{\ignore}[1]{}

\usepackage{commath}
\usepackage{color}
\usepackage{amsmath}
\usepackage{adjustbox}
\usepackage[normalem]{ulem}

\newcommand{\bes} {\begin{subequations}}
\newcommand{\ees} {\end{subequations}}
\newcommand{\beq}{\begin{equation}}
\newcommand{\eeq}{\end{equation}}

\def\>{\rangle}
\def\<{\langle}
\def\Tr{\mathrm{Tr}}

\newcommand{\ketb}[2]{|{#1}\>\!\<#2|}

\def\XY4{\text{XY}4}

\begin{document}

%\title{Probing non-Markovian qubit noise and modeling Post Markovian Master Equation}
\title{Probing Qubit Noise with a Channel-Resolved Post-Markovian Master Equation}

\author{Chun-Tse Li}
\affiliation{Department of Electrical \&  Computer Engineering and Center for Quantum Information Science \& Technology, University of Southern California, Los Angeles, California 90089, USA}
\author{Jingming Tan}
\affiliation{Department of Electrical \&  Computer Engineering and Center for Quantum Information Science \& Technology, University of Southern California, Los Angeles, California 90089, USA}
\author{Vasil Gucev}
\affiliation{Department of Electrical \&  Computer Engineering and Center for Quantum Information Science \& Technology, University of Southern California, Los Angeles, California 90089, USA}
\author{Daniel A. Lidar}
\affiliation{Department of Electrical \&  Computer Engineering and Center for Quantum Information Science \& Technology, University of Southern California, Los Angeles, California 90089, USA}
\affiliation{Departments of Chemistry and Physics \& Astronomy,
University of Southern California, Los Angeles, CA 90089, USA}
\affiliation{Quantum Elements, Inc., Westlake Village, CA}

\begin{abstract}
Accurate noise characterization is essential for scaling quantum processors toward fault-tolerant operation. Although reduced qubit dynamics are often modeled with Markovian master equations, present-day devices can exhibit memory effects generated by residual qubit-qubit couplings, structured environments, and finite bath correlation times. Here we develop a channel-resolved, Post-Markovian Master Equation model for non-Markovian noise and test it in superconducting qubits. Using idle-evolution tomography on IBM Quantum processors, we identify complementary operational signatures of non-Markovianity, including violations of CP-divisibility and revivals of distinguishability-based information-backflow measures. We further derive a closed-form spectator-$ZZ$ model with local dissipation and show that it captures the observed transverse Bloch-vector revivals while leaving the longitudinal relaxation mode Markovian within the model. The fitted closed-form dynamics enable an analytical reconstruction of the transverse memory kernel, whose damped oscillatory structure captures the non-Markovian correction beyond the fitted Markovian baseline. Two-qubit tomography shows buildup and revivals of quantum mutual information on comparable timescales, supporting spectator-induced crosstalk as an important contributor to the observed memory effects. Our results connect operational non-Markovianity diagnostics, microscopic crosstalk modeling, and reduced memory-kernel reconstruction in a single experimental framework for superconducting quantum hardware.
\end{abstract}

\maketitle

\section{Introduction}
\label{sec:intro}

Quantum processors are inherently open quantum systems, and their performance is limited by decoherence, dissipation, residual couplings, and control imperfections associated with uncontrolled degrees of freedom~\cite{alicki_quantum_2007,breuer_theory_2010,rivasvargas_open_2012,lidar_quantum_2013}. While the ideal circuit model assumes perfectly unitary operations, practical quantum hardware must be described in terms of effective noise processes. Accurate reduced noise models are therefore not merely a theoretical concern, but a central requirement for device characterization, calibration, error mitigation, and the long-term development of fault-tolerant quantum computation~\cite{preskill_quantum_2018}.

A standard reduced description of memoryless open-system dynamics is the Gorini-Kossakowski-Lindblad-Sudarshan (GKLS) master equation~\cite{lindblad_generators_1976,gorini_completely_1976}. This approximation successfully captures simple exponential relaxation and dephasing and underlies the common use of coherence benchmarks such as $T_1$ and $T_2$. However, realistic quantum devices often deviate from this picture. In superconducting circuits, temporally correlated noise can arise from structured environments, microscopic defects, residual qubit-qubit couplings, and spectator-induced crosstalk, leading to non-exponential decay, coherence revivals, and information backflow 
\cite{rivas_quantum_2014,breuer_colloquium_2016,devega_dynamics_2017,krinner_benchmarking_2020,jurcevic_effective_2022,tripathi_modeling_2024,sete_error_2024}. In particular, parasitic $ZZ$ interactions are now recognized as a ubiquitous and practically important source of coherent and incoherent crosstalk in superconducting-qubit architectures \cite{mundada_suppression_2019,sarovar_detecting_2020,tripathi_suppression_2022,ni_scalable_2022,cai_impact_2021,zhao_quantum_2022,majumder_realtime_2020}

A number of theoretical frameworks have been developed to describe such non-Markovian dynamics. At the formal level, projection-operator methods lead to exact Nakajima-Zwanzig memory-kernel equations \cite{nakajima_quantum_1958,zwanzig_ensemble_1960}, but these equations are rarely convenient for direct experimental reconstruction. At the opposite end, multitime approaches such as the process tensor provide an operationally complete description of non-Markovian processes and have been experimentally demonstrated on superconducting quantum processors \cite{pollock_nonmarkovian_2018,white_demonstration_2020,white_nonmarkovian_2022}. Between these extremes lie reduced phenomenological models, among which the post-Markovian master equation (PMME) provides one compact way to interpolate between Lindblad evolution and finite-memory corrections \cite{shabani_completely_2005,budini_postmarkovian_2014}. In its standard form, the PMME uses a scalar memory kernel to parameterize the temporal nonlocality of the reduced dynamics, and scalar-kernel variants have been analyzed extensively in the literature. Such approaches have already been shown to capture non-Markovian superconducting-qubit dynamics more accurately than purely Markovian models, and related studies have highlighted the practical relevance of PMME-based descriptions for present-day hardware \cite{zhang_predicting_2022,agarwal_modelling_2024,mazzola_phenomenological_2010,campbell_critical_2012,sutherland_nonmarkovianity_2018}.

Despite this progress, an important gap remains. A scalar-kernel description constrains distinct dynamical sectors to share a common underlying memory function. For realistic superconducting devices, this can be too restrictive: relaxation, dephasing, and spectator-induced crosstalk need not have the same memory timescale or the same physical origin. As a result, a scalar-kernel description can obscure which dynamical sector is genuinely non-Markovian and can introduce fitting artifacts when distinct noise channels are forced into a shared phenomenological form. This limitation becomes especially relevant in superconducting-qubit experiments, where long-time amplitude damping and short-time crosstalk-induced revivals can coexist in the same data set.

In this work, we develop and experimentally test a PMME-inspired, channel-resolved memory-kernel model tailored to superconducting-qubit noise. Rather than describing memory effects through a single scalar kernel, we consider
\begin{equation}
\dot{\rho}(t)
=
\mathcal{L}\rho(t)
+
\int_0^t d\tau\,\ \mathcal{K}(\tau)\rho(t-\tau),
\end{equation}
where $\mathcal{L}$ is the Markovian baseline generator and $\mathcal{K}(\tau)$ is an operator-valued memory kernel. For the single-qubit setting studied here, we use a diagonal-mode ansatz for $\mathcal{K}(\tau)$ in the bi-orthonormal eigenbasis of $\mathcal{L}$, so that distinct dynamical modes can be assigned distinct memory functions. This construction separates long-time Lindblad decay from short-time channel-specific memory effects in a transparent and physically interpretable way, and it recovers scalar-kernel memory models as a constrained case when the mode-dependent kernels are tied to a common function.

Experimentally, we test this framework on IBM superconducting hardware by combining single-qubit and two-qubit tomography with idle-time evolution in the presence of nearby spectator qubits. We diagnose non-Markovianity through complementary operational witnesses, including violations of CP-divisibility and revivals in distinguishability-based measures of information backflow \cite{breuer_measure_2009,rivas_entanglement_2010}. We then use the quantum mutual information between neighboring qubits as an information-theoretic probe of correlation buildup during idle evolution. The observed mutual-information revivals occur on timescales comparable to the single-qubit distinguishability revivals, supporting a crosstalk interpretation. To test a microscopic mechanism for these memory effects, we derive an analytical spectator-$ZZ$ crosstalk model with local dissipation and show that it reproduces the measured single-qubit Bloch-vector dynamics with high accuracy. While spectator-induced $ZZ$ effects and associated dephasing mechanisms have been analyzed previously~\cite{cai_impact_2021,krinner_benchmarking_2020,tripathi_suppression_2022,jurcevic_effective_2022,tripathi_modeling_2024}, we are not aware of a closed-form reduced single-qubit treatment of idle evolution with locally dissipative $ZZ$-coupled spectators that is directly connected to a memory-kernel reconstruction. This closed-form solution enables an analytical reconstruction of the nontrivial channel-resolved memory kernel, providing a compact and physically interpretable summary of device memory that can inform hardware-aware characterization and mitigation strategies.

The structure of this paper is as follows. In \cref{sec:prelim}, we review the CP-divisibility and information-backflow diagnostics, introduce the mutual-information probe of crosstalk, and formulate the channel-resolved memory-kernel model. In \cref{sec:zz-crosstalk-model}, we derive the microscopic spectator-$ZZ$ crosstalk model and its reduced single-qubit dynamics. In \cref{sec:experiment}, we describe the IBM hardware, idle-evolution protocol, and state- and process-tomography procedures. In \cref{sec:results}, we present the CP-divisibility, information-backflow, mutual-information, and memory-kernel reconstruction results. We conclude in \cref{sec:conclusion}.

\section{Preliminaries}
\label{sec:prelim}

In this section, we review the formal definition of non-Markovianity and introduce the information-theoretic quantities used to probe non-Markovian noise on superconducting hardware. We then present the PMME, which provides an analytically tractable description of non-Markovian qubit dynamics beyond the standard Lindblad framework.

\subsection{Non-Markovianity via CP-Divisibility}
\label{sec:cp-div-intro}

A rigorous operational way to characterize non-Markovian dynamics is through CP-divisibility. A one-parameter family of dynamical maps $\{\Phi_t\}_{t\ge0}$ is said to be \emph{CP-divisible} if, for all $t\ge s\ge0$, there exists a completely positive and trace-preserving (CPTP) intermediate map $\mathcal V_{t,s}$ such that
\begin{align}
\Phi_t = \mathcal V_{t,s}\circ \Phi_s .
\end{align}
Within the CP-divisibility definition, the evolution is called \emph{Markovian} when the map is CP-divisible for all $t\ge s$. A lack of CP-divisibility is therefore a witness of non-Markovian behavior in this operational sense.

For any candidate map, complete positivity can be tested through its Choi matrix~\cite{choi_completely_1975}. With the convention
\begin{align}
\label{eq:Choi}
\chi_t = (\mathcal I\otimes \Phi_t)(\ketb{\gamma}{\gamma}),
\end{align}
where
\begin{align}
\ket{\gamma}=\sum_i \ket{i}\otimes\ket{i}
\end{align}
is the unnormalized maximally entangled state on $\mathcal H\otimes\mathcal H$, the map $\Phi_t$ is completely positive if and only if $\chi_t\ge0$. With this same convention, trace preservation is equivalent to $\Tr_2\chi_t=I$, where $\Tr_i$ denotes partial trace of the $i$'th subsystem.

Direct process tomography of the intermediate map $\mathcal V_{t,s}$ is generally infeasible, because it would require preparing arbitrary system states at the intermediate time $s$ while preserving the bath state and system-bath correlations generated during the prior evolution from $0$ to $s$. Instead, we reconstruct the dynamical maps $\Phi_t$ and $\Phi_s$ and infer the intermediate map from their Liouville representations.

With the convention in \cref{eq:Choi}, the inverse Choi-Jamio\l{}kowski map is
\begin{align}
\Phi_t(\rho)
=
\Tr_1\bigl[(\rho^T\otimes I)\chi_t\bigr]
=
\Tr_1\bigl[\chi_t(\rho^T\otimes I)\bigr],
\end{align}
where the transpose is taken in the basis used to define $\ket{\gamma}$. The equality of the two partial-trace expressions follows from cyclicity of the partial trace with respect to operators acting only on the subsystem being traced out. A derivation is given in Appendix~\ref{app:superop-to-choi-matrix}.

Let $\operatorname{vec}$ denote column-stacking vectorization, and label the components of $\operatorname{vec}(\rho)$ by the corresponding matrix indices, so that
\begin{align}
[\operatorname{vec}(\rho)]_{ij}=\rho_{ij}.
\end{align}
The Liouville representation of $\Phi_t$ is then
\begin{align}
\operatorname{vec}\bigl(\Phi_t(\rho)\bigr)=S_t\operatorname{vec}(\rho).
\end{align}
With the Choi convention in \cref{eq:Choi}, the superoperator matrix $S_t$ and the Choi matrix $\chi_t$ are related by the reshuffling rule
\begin{align}
(S_t)_{ij,k\ell}=(\chi_t)_{ki,\ell j}.
\end{align}
Equivalently,
\begin{align}
S_t=\mathcal R(\chi_t),
\end{align}
where $\mathcal R$ is the corresponding index-reshuffling map.

If $S_s$ is invertible, then CP-divisibility implies
\begin{align}
S_t=S_{t,s}S_s,
\end{align}
where $S_{t,s}$ is the Liouville representation of $\mathcal V_{t,s}$. Hence
\begin{align}
S_{t,s}=S_tS_s^{-1}.
\end{align}
The Choi matrix of the intermediate map is therefore
\begin{align}
\chi_{t,s}
=
\mathcal R^{-1}(S_tS_s^{-1}).
\end{align}
The map $\mathcal V_{t,s}$ is CP if and only if $\chi_{t,s}\ge0$. If $\Phi_s$ and $\Phi_t$ are trace preserving and $S_s$ is invertible, then the intermediate map defined above is trace preserving as well. In experimental reconstructions, trace-preservation may not be exact and the residual should be monitored together with the minimum eigenvalue of $\chi_{t,s}$.

If $S_s$ is singular, the inverse formula is not available. In that case, CP-divisibility is the question of whether there exists a CPTP map $\mathcal V_{t,s}$ satisfying
\begin{align}
S_t=S_{t,s}S_s.
\end{align}
Equivalently, in Choi form, one must ask whether there exists a matrix $\chi_{t,s}$ such that
\begin{align}
\mathcal R(\chi_{t,s})S_s=S_t,
\quad
\chi_{t,s}\ge0,
\quad
\Tr_2\chi_{t,s}=I .
\end{align}
Replacing $S_s^{-1}$ by a Moore-Penrose pseudoinverse selects one particular linear extension, but it does not by itself solve this feasibility problem. Therefore, when $S_s$ is singular or ill conditioned, negative eigenvalues obtained from the pseudoinverse construction should be interpreted as a diagnostic rather than as a rigorous proof of the absence of every possible CPTP intermediate map, unless the conclusion is stable under conditioning thresholds and statistical uncertainty.

\subsection{Non-Markovianity via Information Backflow}
\label{sec:info-backflow}

Another operational way to witness non-Markovianity is through information backflow~\cite{breuer_measure_2009,breuer_colloquium_2016}. The basic idea is that, under divisible memoryless evolution, no later dynamical step should increase the distinguishability of two system states. Two useful distinguishability measures are the trace-norm distance and the quantum relative entropy,
\beq
\begin{aligned}
T(\rho,\sigma)
&\coloneqq
\frac{1}{2}\|\rho-\sigma\|_1
=
\frac{1}{2}\Tr|\rho-\sigma|,
\\
D(\rho\Vert\sigma)
&\coloneqq
\begin{cases}
\Tr\big[\rho(\log\rho-\log\sigma)\big],
&
\operatorname{supp}(\rho)\subseteq\operatorname{supp}(\sigma),
\\
+\infty,
&
\text{otherwise},
\end{cases}
\end{aligned}
\eeq
where $|A|\coloneqq\sqrt{A^\dagger A}$. The trace-norm distance has a direct operational meaning in binary state discrimination~\cite{helstrom_quantum_1969}, while the quantum relative entropy is a standard information-theoretic distinguishability~\cite{umegaki_conditional_1962}. Both quantities satisfy a data-processing inequality under CPTP maps~\cite{lindblad_completely_1975}:
\beq
\begin{aligned}
T\big(\Phi(\rho),\Phi(\sigma)\big)
&\le
T(\rho,\sigma),
\\
D\big(\Phi(\rho)\Vert\Phi(\sigma)\big)
&\le
D(\rho\Vert\sigma).
\end{aligned}
\eeq

This contractivity implies monotonic decay in time whenever the dynamics is CP-divisible~\cite{rivas_entanglement_2010,rivas_quantum_2014}. Specifically, if for every $t\ge s\ge0$ the dynamical map can be written as
\beq
\Phi_t=\mathcal V_{t,s}\circ\Phi_s,
\eeq
with $\mathcal V_{t,s}$ CPTP, then the evolution from time $s$ to time $t$ is itself a CPTP processing step. Therefore, for any pair of initial states $\rho_1$ and $\rho_2$, with evolved states
\beq
\rho_j(t)=\Phi_t(\rho_j),
\eeq
one has
\beq
\begin{aligned}
T\big(\rho_1(t),\rho_2(t)\big)
&\le
T\big(\rho_1(s),\rho_2(s)\big),
\\
D\big(\rho_1(t)\Vert\rho_2(t)\big)
&\le
D\big(\rho_1(s)\Vert\rho_2(s)\big),
\quad t\ge s .
\end{aligned}
\eeq

A statistically significant temporary increase of either distinguishability measure is therefore incompatible with CP-divisibility of the reconstructed dynamical family and is commonly interpreted as information flowing back from inaccessible degrees of freedom into the system. In practice, this witness can be implemented by preparing two different initial states, reconstructing their density matrices as a function of time, and monitoring
\beq
T\big(\rho_1(t),\rho_2(t)\big)
\quad
\text{or}
\quad
D\big(\rho_1(t)\Vert\rho_2(t)\big).
\eeq
A revival of either quantity provides evidence of non-Markovian dynamics, while the absence of a revival for a particular pair of probe states does not prove CP-divisibility, since such a conclusion would require control over all state pairs and all time intervals.

When the nominal probe states are orthogonal pure states, the ideal relative entropy is infinite at the initial time because the support condition above is violated. Thus, in the experimental analysis, the relative entropy should be evaluated on full-rank reconstructed density matrices, or after applying a fixed full-rank regularization consistently to both states. The relevant witness is then the occurrence of statistically significant revivals away from such singular points.

\subsection{Information-theoretic probe of crosstalk}
\label{sec:qmi-crosstalk}

An operational signature of crosstalk or correlated noise is the generation of correlations between qubits that were initially prepared in a product state. Let qubits $A$ and $B$ be initialized as $\rho_A\otimes\rho_B$. After joint idle evolution, their state is a generally correlated two-qubit density matrix $\rho_{AB}(t)$, with reduced states
\begin{align}
\rho_A(t)=\Tr_B[\rho_{AB}(t)],
\quad
\rho_B(t)=\Tr_A[\rho_{AB}(t)].
\end{align}
The total correlation in $\rho_{AB}(t)$ can be quantified by the quantum mutual information~\cite{nielsen_quantum_2010,groisman_quantum_2005},
\begin{align}
I(A:B)_{\rho(t)}
=
S\bigl(\rho_A(t)\bigr)
+
S\bigl(\rho_B(t)\bigr)
-
S\bigl(\rho_{AB}(t)\bigr),
\end{align}
where
\begin{align}
S(\rho)=-\Tr[\rho\log_2\rho]
\end{align}
is the von Neumann entropy in bits. Equivalently,
\begin{align}
I(A:B)_{\rho(t)}
=
D\bigl(\rho_{AB}(t)\big\Vert\rho_A(t)\otimes\rho_B(t)\bigr),
\end{align}
so $I(A:B)_{\rho(t)}\ge0$, with equality if and only if
\begin{align}
\rho_{AB}(t)=\rho_A(t)\otimes\rho_B(t).
\end{align}
For two qubits, this normalization gives
\begin{align}
0\le I(A:B)_{\rho(t)}\le2 .
\end{align}

Thus, growth of $I(A:B)_{\rho(t)}$ from an initially product state witnesses the buildup of correlations between the qubits. These correlations can be classical or quantum, and the mutual information is not by itself an entanglement witness. Moreover, a positive mutual information does not uniquely identify a microscopic mechanism: it can arise from coherent qubit-qubit coupling, correlated noise, common-mode environmental fluctuations, or residual state-preparation and measurement correlations. In the present setting, we therefore use $I(A:B)_{\rho(t)}$ as an information-theoretic probe of correlation buildup that is consistent with crosstalk, rather than as a standalone proof of its physical origin~\cite{sarovar_detecting_2020,zhao_quantum_2022}.

In practice, we tomographically reconstruct $\rho_{AB}(t)$ at discrete idle times and track $I(A:B)_{\rho(t)}$. Since imperfect state preparation and readout can produce a small nonzero baseline value at the earliest time point, the relevant diagnostic is the time-dependent growth and revival of $I(A:B)_{\rho(t)}$ relative to this baseline. Correlation revivals that occur on the same timescales as single-qubit distinguishability revivals provide supporting evidence that the observed memory effects are associated with qubit-qubit crosstalk or other correlated device-level dynamics.

\subsection{Channel-resolved post-Markovian memory-kernel model}
\label{sec:PMME}

We start from a single-qubit GKLS generator that captures the baseline Markovian dynamics~\cite{gorini_completely_1976,lindblad_generators_1976}:
\beq
  \label{eq:Lindblad}
  \mathcal{L}(\rho)
    =
      -i[H,\rho] + \gamma_{\downarrow}\mathcal D_{\sigma^-}(\rho) +  \gamma_{\phi}\mathcal D_{Z}(\rho)
\eeq
where $H = -\frac{1}{2}\omega_0Z$, the dissipator is
\begin{align}
\label{eq:dissipator}
    \mathcal D_L(\rho)
    =
    L\rho L^\dagger
    -
    \frac{1}{2}\{L^\dagger L,\rho\},
\end{align}
$Z=\ketb{0}{0}-\ketb{1}{1}$, $\sigma^-=\ketb{0}{1}$, and $\sigma^+=(\sigma^-)^\dagger$. Here $\omega_0$ is the qubit frequency in the chosen frame, $\gamma_{\downarrow}$ is the amplitude-damping rate, and $\gamma_{\phi}$ is the coefficient of the pure-dephasing term. With this convention, the pure-dephasing term contributes $2\gamma_{\phi}$ to the transverse decay rate.

It is convenient to diagonalize \cref{eq:Lindblad} using right eigenoperators $\{R_i\}$ and dual left eigenoperators $\{L_i\}$ of $\mathcal{L}$ such that $\Tr\left[L_iR_j\right]=\delta_{ij}$. We refer to the eigenbasis index values $i$ as \emph{modes}, and define the \emph{damped basis} by
\begin{align}
\mathcal{L}(R_i)=\lambda_iR_i,
\quad
\Tr\left[L_i\mathcal{L}(X)\right]
=
\lambda_i\Tr\left[L_iX\right],
\end{align}
for all single-qubit operators $X$. For \cref{eq:Lindblad}, one obtains
\begin{align}
\label{eq:RLlambda}
\begin{array}{lll}
  R_0=\dfrac{I+Z}{\sqrt2},
  &
  L_0=\dfrac{I}{\sqrt2},
  &
  \lambda_0=0,
  \\[0.35cm]
  R_1=\dfrac{Z}{\sqrt2},
  &
  L_1=\dfrac{Z-I}{\sqrt2},
  &
  \lambda_1=-\gamma_{\downarrow},
  \\[0.35cm]
  R_2=\sigma^-,
  &
  L_2=\sigma^+,
  &
  \lambda_2=i\omega_0-\left(2\gamma_{\phi}+\frac{\gamma_{\downarrow}}{2}\right),
  \\[0.35cm]
  R_3=\sigma^+,
  &
  L_3=\sigma^-,
  &
  \lambda_3=\lambda_2^*.
\end{array}
\end{align}
Thus $R_0$ is the stationary fixed point of the amplitude-damping dynamics, $R_1$ is the longitudinal relaxation mode, and $R_2$ and $R_3$ are the two transverse coherence modes.

To capture structured memory effects beyond this Markovian baseline, we introduce a channel-resolved memory kernel~\cite{breuer_theory_2010,shabani_completely_2005,budini_postmarkovian_2014}. The reduced dynamics is modeled by the integro-differential equation
\begin{align}
\dot{\rho}(t)
=
\mathcal{L}\rho(t)
+
\int_0^t d\tau\,
\mathcal{K}(\tau)\rho(t-\tau).
\label{eq:channel-memory-master-equation}
\end{align}
Here $\mathcal{K}(\tau)$ is an operator-valued memory kernel that parameterizes deviations from the baseline GKLS dynamics. We use a diagonal-mode ansatz in the damped basis,
\begin{align}
\mathcal{K}(\tau)X
=
\sum_{i=0}^3
k_i(\tau)R_i\Tr\left(L_iX\right),
\label{eq:diagonal-kernel-ansatz}
\end{align}
where $k_i(\tau)$ is the scalar memory function associated with the $i$th dynamical mode. This ansatz should be understood as a reconstruction model: arbitrary choices of the functions $k_i(\tau)$ need not generate CPTP evolution. In our work, the kernels are inferred from physical tomographic data and from the fitted microscopic spectator-crosstalk model.

Expanding the density matrix as
\begin{align}
\label{eq:rho-expand}
\rho(t)=\sum_{i=0}^3\mu_i(t)R_i,
\quad
\mu_i(t)=\Tr\left[L_i\rho(t)\right],
\end{align}
and substituting \cref{eq:diagonal-kernel-ansatz} into \cref{eq:channel-memory-master-equation} yields the decoupled scalar equations
\begin{align}
\dot{\mu}_i(t)
=
\lambda_i\mu_i(t)
+
\int_0^t d\tau\,
k_i(\tau)\mu_i(t-\tau).
\label{eq:mode-memory-equation}
\end{align}
Let
\begin{align}
\tilde f(z)=\int_0^\infty dt e^{-zt}f(t)
\end{align}
denote the Laplace transform. Taking the Laplace transform of \cref{eq:mode-memory-equation} gives
\begin{align}
\label{eq:mu-laplace}
\tilde{\mu}_i(z)
=
\frac{\mu_i(0)}{z-\lambda_i-\tilde{k}_i(z)}.
\end{align}
For modes with $\mu_i(0)\ne0$, define the normalized mode function
\begin{align}
\label{eq:xi}
\xi_i(t)
\coloneqq
\frac{\mu_i(t)}{\mu_i(0)},
\quad
\tilde{\xi}_i(z)
=
\frac{\tilde{\mu}_i(z)}{\mu_i(0)}.
\end{align}
Then \cref{eq:mu-laplace} implies
\beq
\begin{aligned}
\label{eq:kernel-s-domain}
\tilde{k}_i(z)
&=
z-\lambda_i-\frac{1}{\tilde{\xi}_i(z)}\\
k_i(t)
&=
\operatorname{Lap}^{-1}_{z\to t}
\left[
z-\lambda_i-\frac{1}{\tilde{\xi}_i(z)}
\right].
\end{aligned}
\eeq
\Cref{eq:kernel-s-domain} is the basic mode-resolved reconstruction formula: once the dynamical coefficients $\mu_i(t)$ are obtained, the corresponding memory kernel $k_i(t)$ can be extracted mode by mode after subtracting the baseline GKLS contribution $\lambda_i$. This is also the key difference between our approach and the original PMME construction~\cite{shabani_completely_2005}, which contained only a single memory kernel $k(t)$.

Trace preservation fixes the identity sector. Since $R_0$ is the only right eigenoperator with nonzero trace, and $L_0=I/\sqrt2$, a trace-one state satisfies
\begin{align}
\mu_0(t)=\Tr[\rho(t)]/\sqrt2=1/\sqrt2
\end{align}
for all $t$. Hence
\begin{align}
\xi_0(t)=1,
\quad
\tilde{\xi}_0(z)=\frac{1}{z},
\end{align}
and since $\lambda_0=0$,
\begin{align}
\tilde{k}_0(z)
=
z-\lambda_0-\frac{1}{\tilde{\xi}_0(z)}
=
0.
\end{align}
Thus the trace sector carries no memory kernel.

In the microscopic spectator-$ZZ$ model derived in \cref{sec:zz-crosstalk-model} below, the longitudinal mode is also purely Markovian: the spectator coupling modifies the transverse coherences but does not alter the local amplitude-damping equation for the main qubit. Within this model,
\begin{align}
k_1(t)=0.
\end{align}
For the transverse modes, Hermiticity of $\rho(t)$ implies
\begin{align}
\mu_3(t)=\mu_2(t)^*,
\quad
\lambda_3=\lambda_2^*,
\quad
k_3(t)=k_2(t)^*,
\end{align}
as derived in Appendix~\ref{app:complex-conjugate}. Therefore, for the spectator-$ZZ$ dynamics considered here, the only nontrivial memory contribution is carried by the transverse kernel $k_2(t)$, which encodes the coherence revivals induced by spectator-qubit crosstalk.

In practice, reconstructing $k_2(t)$ directly from \cref{eq:kernel-s-domain} by applying a numerical Bromwich inversion to finite, noisy tomographic data is ill conditioned. To avoid this instability, in the following section we derive a closed-form expression for the reduced single-qubit dynamics generated by spectator-induced $ZZ$ crosstalk with local dissipation. This gives an analytical form for the transverse mode function $\xi_2(t)$. We then fit the parameters of this expression to the experimental Bloch-vector data and substitute the fitted $\xi_2(t)$ into \cref{eq:kernel-s-domain}. Because the fitted mode function is available in closed form, the inverse Laplace transform can be evaluated analytically, using partial-fraction methods, to obtain the corresponding channel-resolved memory kernel within the fitted model.

\section{Microscopic spectator-$ZZ$ crosstalk model}
\label{sec:zz-crosstalk-model}

To test a microscopic mechanism underlying the reconstructed memory kernel, we model the main qubit as coupled to nearby spectator qubits through parasitic $ZZ$ interactions, while all qubits undergo local Markovian dissipation. This model is physically motivated in superconducting-qubit architectures, where residual $ZZ$ couplings are a common source of coherent spectator-induced errors and crosstalk~\cite{mundada_suppression_2019,sarovar_detecting_2020,tripathi_suppression_2022,ni_scalable_2022,cai_impact_2021,zhao_quantum_2022,majumder_realtime_2020}. On the heavy-hex devices used in this work, a qubit has at most three neighboring spectators, so we consider a main qubit, labeled $0$, coupled to $N\le3$ surrounding qubits. The joint main-spectator dynamics is generated by
\begin{align}
\mathcal{L}_{\mathrm{tot}}(\rho)
=
-i[H,\rho]
+
\sum_{q=0}^{N}\left(\gamma_{\downarrow,q}\mathcal D_{\sigma_q^-}(\rho) + \gamma_{\phi,q}\mathcal D_{Z_q}(\rho)\right),
\end{align}
with
\begin{align}
H=
-\frac{1}{2}\omega_0 Z_0
+
\sum_{q=1}^{N}
\frac{1}{2}J_{0q}Z_0Z_q .
\end{align}
Here $\omega_0$ is the main-qubit frequency in the chosen rotating frame, and $J_{0q}$ is the residual $ZZ$ coupling between the main qubit and spectator $q$. The enlarged main-spectator dynamics is Markovian by construction; non-Markovianity appears in the reduced dynamics of the main qubit after the spectator degrees of freedom are traced out.

An important consequence of the pure $ZZ$ form of the coupling is that it affects only the transverse coherence sector of the main qubit, while the longitudinal mode remains Markovian. Since
\begin{align}
[Z_0,H]=0,
\end{align}
the spectator coupling does not enter the Hamiltonian equation of motion for the longitudinal operator $Z_0$. The spectator dissipators act only on the spectator qubits, and the main-qubit pure-dephasing term also leaves $Z_0$ invariant. Thus the reduced longitudinal mode is governed only by main-qubit amplitude damping:
\begin{align}
\dot{\mu}_1(t)=\lambda_1\mu_1(t),
\quad
\lambda_1=-\gamma_{\downarrow,0}.
\end{align}
For an initial state with $\mu_1(0)\ne0$, this gives
\begin{align}
\xi_1(t)=e^{\lambda_1 t},
\quad
\tilde{\xi}_1(z)=\frac{1}{z-\lambda_1}.
\end{align}
Substituting into \cref{eq:kernel-s-domain} yields
\begin{align}
\tilde{k}_1(z)
=
z-\lambda_1-\frac{1}{\tilde{\xi}_1(z)}
=
0,
\quad
k_1(t)=0.
\end{align}
Thus, within the spectator-$ZZ$ crosstalk model, the nontrivial memory contribution is carried by the transverse modes, whose dynamics inherit the oscillatory structure generated by the spectator coupling.

For the two transverse modes, we derive the closed-form Bloch-vector dynamics in Appendix~\ref{app:ZZ-crosstalk}. In the single-spectator case, with the spectator initialized in $\ket{+}$, the normalized transverse mode is
\begin{align}
\xi_2(t)
=
e^{-\left(\Gamma_0+\frac{\Gamma_1}{2}-i\omega_0\right)t}
\left[
\cosh(\Omega_1 t)
+
\frac{\Gamma_1}{2\Omega_1}\sinh(\Omega_1 t)
\right],
\end{align}
where
\begin{align}
\Gamma_0
=
\frac{\gamma_{\downarrow,0}}{2}
+
2\gamma_{\phi,0},
\quad
\Gamma_1=\gamma_{\downarrow,1},
\quad
\Omega_1
=
\frac{\Gamma_1}{2}
-
iJ_{01}.
\end{align}
In the limit $J_{01}\to0$, the spectator modulation becomes trivial and
\begin{align}
\xi_2(t)
=
e^{\lambda_2 t},
\quad
\lambda_2
=
-\Gamma_0+i\omega_0,
\end{align}
as expected for the baseline single-qubit GKLS evolution. Thus the residual $ZZ$ coupling generates a non-Markovian modulation on top of the baseline Lindblad decay.

More generally, for $N\le3$ independent spectators prepared with zero initial $Z$ polarization, such as $\ket{+}^{\otimes N}$, $\ket{-}^{\otimes N}$, $\ket{+i}^{\otimes N}$, or $\ket{-i}^{\otimes N}$, the transverse coherence factorizes into independent spectator modulations:
\begin{align}
\label{eq:xi2-1}
\xi_2(t)
=
e^{\lambda_2 t}
\prod_{q=1}^{N}\Phi_q(t),
\end{align}
where
\beq
\label{eq:xi2-2}
\begin{aligned}
\Phi_q(t)
&=
e^{-\frac{\Gamma_q}{2}t}
\left[
\cosh(\Omega_q t)
+
\frac{\Gamma_q}{2\Omega_q}\sinh(\Omega_q t)
\right]\\
\Omega_q
&=
\frac{\Gamma_q}{2}
-
iJ_{0q},
\end{aligned}
\eeq
and $\Gamma_q=\gamma_{\downarrow,q}$ for $q\ge1$. Local pure dephasing of the spectator qubits does not enter this reduced main-qubit coherence factor, because the coupling is diagonal in the spectator $Z_q$ basis and spectator dephasing does not change the spectator $Z_q$ populations. More general product spectator preparations, including nonzero initial $Z$ polarization, are treated in Appendix~\ref{app:ZZ-crosstalk}. The expressions above provide the fitted mode function $\xi_2(t)$ we use to reconstruct the transverse memory kernel.

\section{Experiment Setup}
\label{sec:experiment}

\begin{figure}[t]
    \centering
    \includegraphics[width=0.45\linewidth]{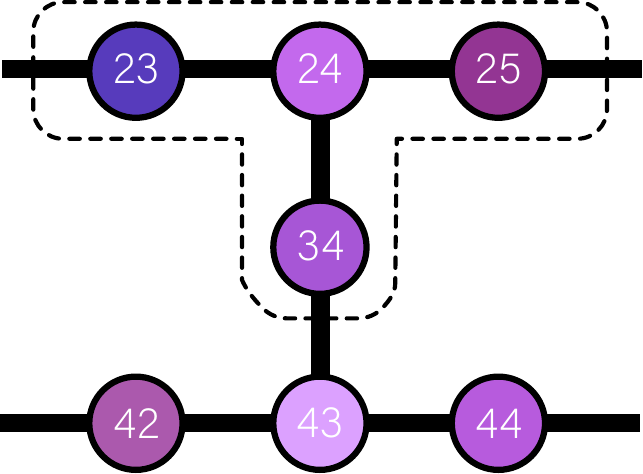}
    \caption{\justifying
    Layout of the IBM heavy-hex superconducting processor used in the experiment, shown as a partial device view. We select qubits 23, 24, 25, and 34, with qubit 24 serving as the main qubit and the other three qubits acting as nearby spectators.
    }
    \label{fig:connectivity}
\end{figure}

\begin{figure*}[t!]
    \centering
    \includegraphics[width=0.9\linewidth]{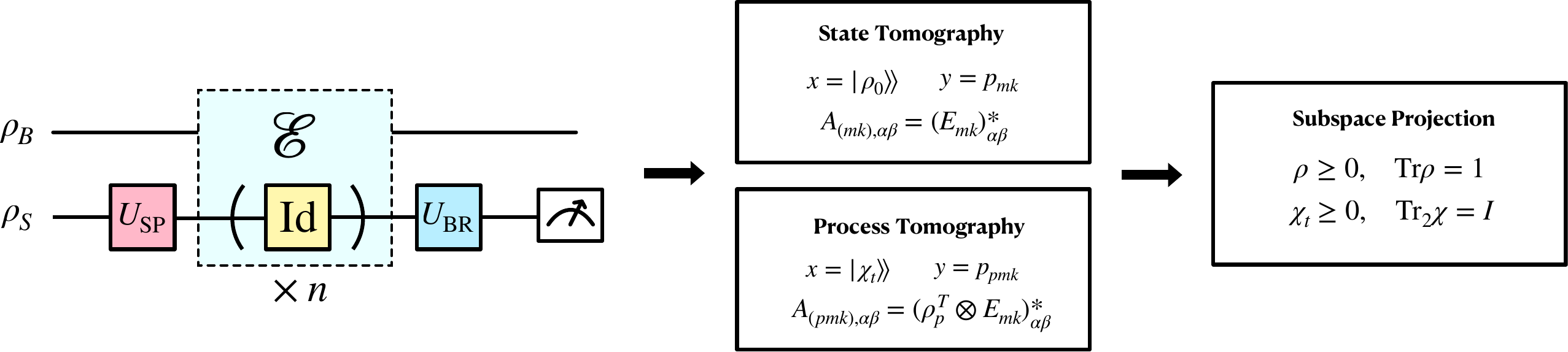}
    \caption{\justifying
    Schematic of the experimental workflow. The circuit consists of three stages: \textbf{(1)} state preparation, implemented by $U_{\mathrm{SP}}$; \textbf{(2)} idle evolution, implemented by a variable number of identity gates; and \textbf{(3)} basis rotations, implemented by $U_{\mathrm{BR}}$, followed by computational-basis measurement. By varying the number of identity gates, we vary the idle time during which the main qubit interacts with the spectator qubits and other environmental degrees of freedom. The measurement statistics are assembled into a probability vector and combined with the corresponding measurement effects to perform either state tomography or process tomography. The resulting raw estimates are projected onto the physical subspace, enforcing $\rho\ge 0$ and $\Tr\rho=1$ for states, and $\chi_t\ge 0$ and $\Tr_2\chi_t=I$ for Choi matrices in the convention of \cref{eq:Choi}.
    }
    \label{fig:setup}
\end{figure*}

In this section, we describe the experimental setup used to probe the non-Markovian noise processes introduced above. We select a four-qubit subsystem on an IBM superconducting processor, consisting of one ``main'' qubit whose reduced dynamics we monitor and three neighboring spectator qubits that act as an engineered local environment. The selected connectivity is shown in \cref{fig:connectivity}. By initializing the spectator qubits in prescribed product states, we probe how spectator-induced correlations affect the evolution of the main qubit. A schematic of the tomography workflow is shown in \cref{fig:setup}.

\subsection{Quantum State Tomography}
\label{sec:state-tomography}

To track the time-dependent reduced density matrix of the main qubit, we perform quantum state tomography at each probe time $t$~\cite{james_measurement_2001,paris_quantum_2004,blume-kohout_optimal_2010}. Let
\begin{align}
\rho_0(t)=\Tr_{\mathrm{spec}}\rho_{\mathrm{joint}}(t)
\end{align}
denote the reduced state of the main qubit after tracing out the spectator qubits. For single-qubit tomography, we measure the main qubit in the Pauli bases
\begin{align}
m\in\{X,Y,Z\},
\end{align}
with binary outcomes $k\in\{0,1\}$. The corresponding measurement effects are
\begin{align}
E_{mk}=U_m^\dagger\ketb{k}{k}U_m,
\end{align}
where $U_m$ is the basis-rotation gate for measurement in basis $m$. The ideal probability for outcome $k$ in basis $m$ is
\begin{align}
p_{mk}(t)=\Tr\left[E_{mk}\rho_0(t)\right].
\end{align}
Each measurement circuit is repeated with $N_{\mathrm{shots}}$ shots, giving an empirical estimate $\hat p_{mk}(t)$ of $p_{mk}(t)$.

For each time $t$, we stack the empirical probabilities $\hat p_{mk}(t)$ into a data vector $y(t)$ and write the density matrix as a vector $x(t)=\operatorname{vec}[\rho_0(t)]$, using column-stacking vectorization. The reconstruction problem is then the linear system
\begin{align}
A x(t)\approx y(t),
\end{align}
with matrix elements
\begin{align}
A_{(mk),\alpha\beta}
=
\left(E_{mk}\right)^*_{\alpha\beta}.
\end{align}
With the Hilbert-Schmidt convention
\begin{align}
\label{eq:HS-IP}
\langle\!\langle A|B\rangle\!\rangle
=
\Tr(A^\dagger B),
\end{align}
equivalently, each row of $A$ is the Hilbert-Schmidt dual vector $\langle\!\langle E_{mk}|$, so that
\begin{align}
(Ax)_{mk}
=
\langle\!\langle E_{mk}|\rho_0\rangle\!\rangle
=
\Tr\left[E_{mk}\rho_0\right].
\end{align}
Because the Pauli-basis measurements form an overcomplete tomographic set, the system is solved in the least-squares sense. The linear-inversion estimate is
\begin{align}
x_{\mathrm{lin}}(t)=A^{+}y(t),
\end{align}
where $A^{+}$ denotes the Moore-Penrose pseudoinverse, and $x_{\mathrm{lin}}(t)$ is reshaped into a raw matrix $\rho_{\mathrm{lin}}(t)$.

Finite-shot noise and calibration errors can make $\rho_{\mathrm{lin}}(t)$ non-Hermitian, non-positive, or slightly non-unit-trace. We therefore enforce physicality before using the reconstructed state in the information-backflow and memory-kernel analyses. A simple projection procedure is to symmetrize $\rho_{\mathrm{lin}}(t)$, set negative eigenvalues to zero, and renormalize the trace. Equivalently, one can compute the physical estimate by solving the constrained least-squares problem
\begin{align}
\hat\rho_0(t)
=
\arg\min_{\rho=\rho^\dagger,\rho\ge 0,\Tr\rho=1}
\sum_{mk}
\left[
\Tr\left(E_{mk}\rho\right)-\hat p_{mk}(t)
\right]^2.
\end{align}
The resulting $\hat\rho_0(t)$ is the physical single-qubit density matrix used in the subsequent analysis.

\subsection{Quantum Process Tomography}
\label{sec:process-tomography}

Quantum process tomography reconstructs an unknown quantum channel rather than a fixed quantum state~\cite{poyatos_complete_1997,chuang_prescription_1997,mohseni_quantumprocess_2008}. In our setting, the reconstructed object is the main-qubit map at idle time $t$, with the spectator preparation held fixed. More explicitly, for a fixed spectator state $\rho_{\mathrm{spec}}$, the reduced map is
\begin{align}
\mathcal E_t^{(\rho_{\mathrm{spec}})}(\rho)
=
\Tr_{\mathrm{spec}}
\left[
\mathcal E_t^{\mathrm{joint}}
\left(
\rho\otimes\rho_{\mathrm{spec}}
\right)
\right].
\end{align}
This product-state initialization, with the same spectator state used for every input probe state, is required for the reconstructed object to be interpreted as a linear CPTP map on the main qubit. In what follows we suppress the spectator-state superscript and write this reduced map simply as $\mathcal E_t$.

To perform process tomography, we prepare a tomographically complete set of input states $\{\rho_p\}$. For single-qubit process tomography, one convenient choice is
\begin{align}
\rho_p = \ketb{p}{p},\quad p\in\left\{0,1,+,+i\right\}.
\end{align}
For each input state, we let the system idle for time $t$ and then measure the output in a tomographically complete set of measurement effects $\{E_{mk}\}$. The ideal measurement probability is
\begin{align}
p_{pmk}(t)
=
\Tr\left[
E_{mk}\mathcal E_t(\rho_p)
\right].
\end{align}
Using the Choi convention in \cref{eq:Choi}, this probability can be written as
\begin{align}
p_{pmk}(t)
=
\Tr\left[
\chi_t
\left(
\rho_p^T\otimes E_{mk}
\right)
\right],
\label{eq:qpt-probability-choi}
\end{align}
where the transpose is taken in the basis used to define $\ket{\gamma}$.

For linear inversion, define
\begin{align}
M_{pmk}
=
\rho_p^T\otimes E_{mk}.
\end{align}
We stack the empirical probabilities $\hat p_{pmk}(t)$ into a vector $y(t)$ and define
\begin{align}
x(t)=\operatorname{vec}(\chi_t).
\end{align}
With the Hilbert-Schmidt inner product [\cref{eq:HS-IP}], 
\cref{eq:qpt-probability-choi} becomes
\begin{align}
p_{pmk}(t)
=
\langle\!\langle M_{pmk}|\chi_t\rangle\!\rangle,
\end{align}
since $M_{pmk}$ is Hermitian. Thus the tomography problem can be written as the linear system
\begin{align}
A x(t)\approx y(t),
\end{align}
where each row of $A$ is the functional
\begin{align}
A_{(pmk),:}
=
\langle\!\langle M_{pmk}|.
\end{align}
Equivalently, under column-stacking vectorization,
\begin{align}
p_{pmk}(t)
=
\operatorname{vec}\left(M_{pmk}^T\right)^T
\operatorname{vec}(\chi_t).
\end{align}

The raw linear-inversion estimate is
\begin{align}
x_{\mathrm{lin}}(t)=A^+y(t),
\end{align}
where $A^+$ denotes the Moore-Penrose pseudoinverse. Reshaping $x_{\mathrm{lin}}(t)$ gives a raw Choi matrix $\chi_{\mathrm{lin}}(t)$. Because finite-shot noise and calibration errors can make $\chi_{\mathrm{lin}}(t)$ non-Hermitian, non-positive, or slightly non-trace-preserving, we enforce physicality before using the reconstructed channel in the CP-divisibility analysis. As noted above, with the Choi convention in \cref{eq:Choi}, a CPTP map satisfies
\begin{align}
\mathcal C \coloneqq\left\{\chi=\chi^\dagger,
\chi\ge 0,
\Tr_2\chi=I\right\} .
\end{align}
A physical estimate can therefore be obtained by solving the constrained least-squares problem
\begin{align}
\hat\chi_t
=
\arg\min_{\chi\in\mathcal C}
\sum_{pmk}
\left(
\Tr\left[
\chi
\left(
\rho_p^T\otimes E_{mk}
\right)
\right]
-
\hat p_{pmk}(t)
\right)^2.
\end{align}
The resulting $\hat\chi_t$ is the CPTP Choi matrix used to construct the Liouville superoperator $S_t=\mathcal R(\hat\chi_t)$ for the subsequent CP-divisibility test. 
In practice, we used Qiskit's process tomography implementation to obtain the physical Choi matrix. We used its constrained least-squares option rather than eigenvalue clipping: the latter enforces complete positivity but does not, in general, preserve the trace-preservation constraint, whereas the constrained least-squares reconstruction enforces both $\chi_t\ge 0$ and $\Tr_2\chi_t=I$.

\section{Results}
\label{sec:results}

In this section we apply a suite of operational diagnostics to the experimentally reconstructed states and dynamical maps. These diagnostics probe both memory-induced non-Markovian effects and correlation buildup between nearby qubits. Specifically, we (i) test CP-divisibility through the Choi matrix of the inferred intermediate map, (ii) examine information backflow through trace-norm distance and quantum relative entropy, (iii) probe qubit-qubit correlations through two-qubit quantum mutual information, and (iv) reconstruct the channel-resolved memory kernel.

\subsection{CP-divisibility test}
\label{sec:results-cp-div}

\begin{figure*}[t!]
     \centering
     \begin{subfigure}[b]{0.45\textwidth}
         \centering
         \includegraphics[width=\textwidth]{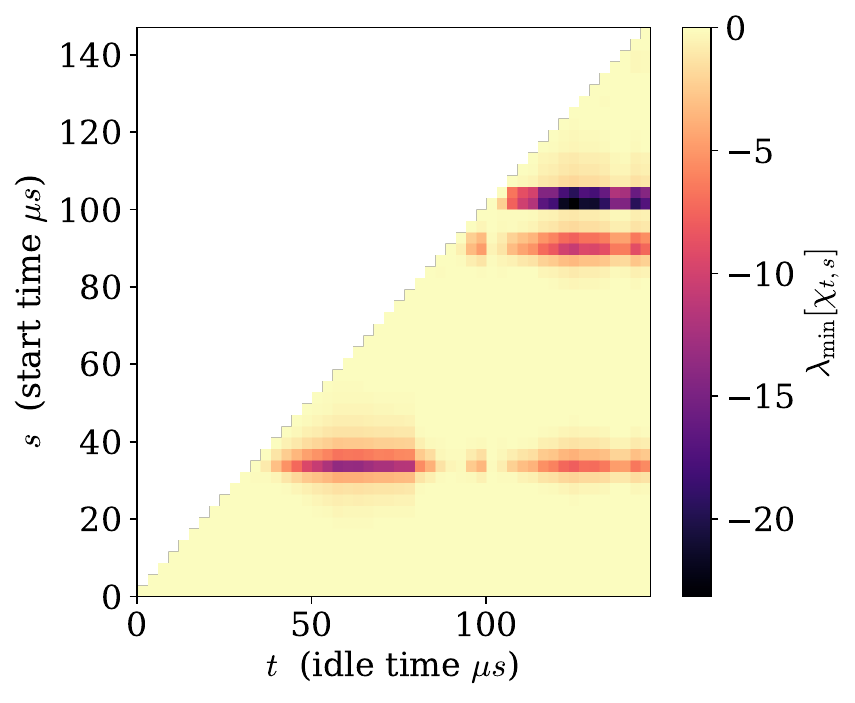}
     \end{subfigure}
     \hfill
     \begin{subfigure}[b]{0.45\textwidth}
         \centering
         \includegraphics[width=\textwidth]{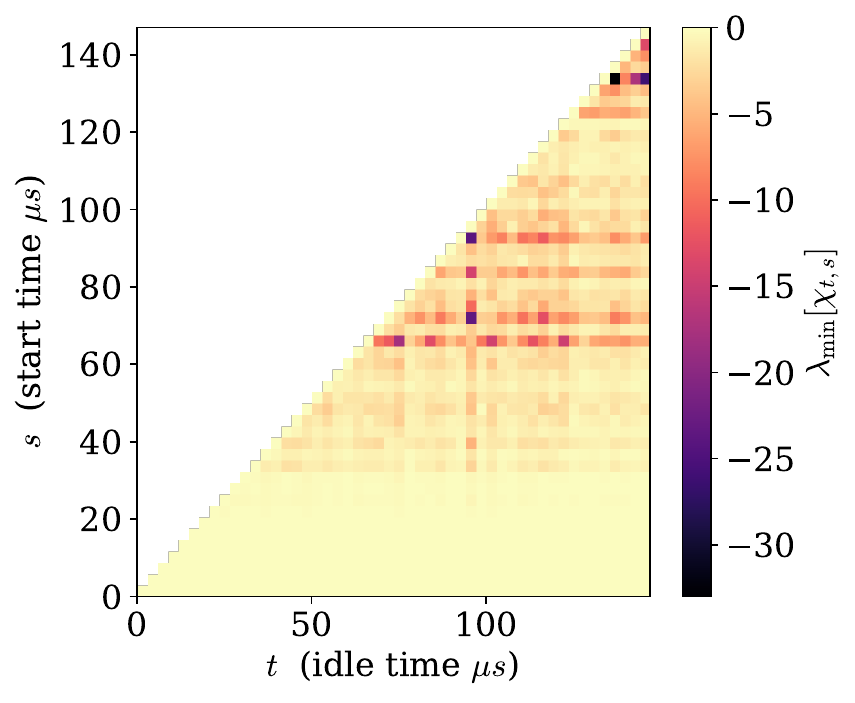}
     \end{subfigure}
     \caption{\justifying
     Heat maps of the minimum eigenvalue $\lambda_{\min}[\chi_{t,s}]$ of the inferred intermediate Choi matrix, plotted as a function of start time $s$ and final idle time $t$, measured on IBM's 127-qubit ``Strasbourg'' processor.
    \textbf{(a)} Main qubit 4 with spectator qubits $\{3,5,15\}$.
    \textbf{(b)} Main qubit 25 with spectator qubits $\{26,27,16\}$.
    The spectator qubits are initialized in $\ket{+}^{\otimes3}$. For each pair $0\le s\le t$, process tomography reconstructs the maps at idle times $s$ and $t$, and the intermediate map is inferred from their Liouville representations. The triangular region $s>t$ is masked. Light-yellow regions have $\lambda_{\min}[\chi_{t,s}]\approx0$, while orange to purple regions have negative minimum eigenvalues, indicating violations of complete positivity for the reconstructed intermediate map.
    }
    \label{fig:cp-div}
\end{figure*}

As discussed in \cref{sec:cp-div-intro}, a dynamical map $\Phi_t$ is CP-divisible if, for every $0\le s\le t$, there exists a CPTP intermediate map $\mathcal V_{t,s}$ such that
$\Phi_t=\mathcal V_{t,s}\circ\Phi_s$. If $\Phi_s$ is invertible, then the corresponding Liouville superoperator is
$S_{t,s}=S_tS_s^{-1}$,
and the Choi matrix of the intermediate map is
$\chi_{t,s} = \mathcal R^{-1}(S_tS_s^{-1})$,
where $\mathcal R$ is the reshuffling map between Choi and Liouville representations. The map $\mathcal V_{t,s}$ is CP if and only if all eigenvalues of $\chi_{t,s}$ are non-negative.

Experimentally, we proceed as follows:
\begin{enumerate}[leftmargin=*]
  \item We reconstruct the Choi matrix $\chi_t$ at each idle time $t$ by process tomography.
  \item We reshuffle $\chi_t$ into the Liouville superoperator $S_t=\mathcal R(\chi_t)$.
  \item For well-conditioned $S_s$, we compute $S_{t,s}=S_tS_s^{-1}$. When $S_s$ is singular or ill conditioned, we replace $S_s^{-1}$ by a Moore-Penrose pseudoinverse $S_s^+$ and regard the resulting matrix
  \begin{align}
  S_{t,s}^{(+)}=S_tS_s^+
  \end{align}
  as a diagnostic linear extension rather than as the unique physical intermediate map.
  \item We compute the corresponding diagnostic Choi matrix
  \begin{align}
  \chi_{t,s}^{(+)}
  =
  \mathcal R^{-1}(S_{t,s}^{(+)})
  \end{align}
  and evaluate its minimum eigenvalue $\lambda_{\min}$.
\end{enumerate}

\Cref{fig:cp-div} shows triangular heat maps of $\lambda_{\min}[\chi_{t,s}]$ over the domain $0\le s\le t\le T$. The light-yellow regions, where $\lambda_{\min}\approx0$, are consistent with CP intermediate dynamics within the reconstruction resolution. In contrast, extended orange to purple regions have negative minimum eigenvalues and therefore indicate violations of CP-divisibility for the inferred intermediate maps. Because pseudoinverse-based intermediate maps can be sensitive to conditioning, we interpret these negative regions as a diagnostic of non-Markovian memory effects. In our analysis the pseudoinverse $S_s^+$ was computed using \texttt{numpy.linalg.pinv}. This routine uses an SVD-based pseudoinverse: singular values smaller than $10^{-15}\sigma_{\max}$, where $\sigma_{\max}$ is the largest singular value of $S_s$, are treated as zero and are not inverted.
% \DL{Is this something you tested, and do they?}

% {\tt \color{blue} Chun-Tse: Yes I think so, I directly used numpy's pseudo-inverse for this. And I think there is a default conditioning threshold $10^{-15}$.} \DL{Then this is the procedure that should be explained, including mentioning that numpy's pseudo-inverse was used.}

% {\tt \color{blue} Chun-Tse: actually, here I applied the pseudo-inverse with singular value threshold $10^{-15}$ to the Choi matrix obtained from the constrained optimization problem. I'm not sure how we can consider the statistical uncertainty here since clipping the singular value below the statistical uncertainty might let the Choi matrix become non TP.
% }

\subsection{Information backflow}
\label{sec:results-info-backflow}

A second signature of non-Markovianity is the backflow of information from inaccessible degrees of freedom to the system. As discussed in \cref{sec:info-backflow}, both trace-norm distance and quantum relative entropy are contractive under CPTP maps. A statistically significant temporary increase in either quantity is incompatible with CP-divisibility and provides evidence of information backflow.

In our experiments, we consider two orthogonal input pairs for the main qubit:
\begin{align}
\begin{array}{lll}
  &\rho_1(0)=\ketb{+}{+},
  &\rho_2(0)=\ketb{-}{-},
  \\[0.25cm]
  &\rho_1(0)=\ketb{+i}{+i},
  &\rho_2(0)=\ketb{-i}{-i}.
\end{array}
\end{align}
The spectator qubits are initialized either in $\ket{+}^{\otimes3}$ or in $\ket{1}^{\otimes3}$. At each idle time $t$, we perform full state tomography on the main qubit, reconstruct $\rho_1(t)$ and $\rho_2(t)$, and compute the trace-norm distance
$T\bigl(\rho_1(t),\rho_2(t)\bigr)$.

For the relative entropy, the ideal orthogonal initial pairs have disjoint support and hence infinite relative entropy at $t=0$. We therefore evaluate $D\bigl(\rho_1(t)\Vert\rho_2(t)\bigr)$ only on full-rank reconstructed density matrices; when numerical eigenvalues vanish, we use the same small full-rank regularization for both states before computing $D$ to ensure that the matrix logarithm is well defined.

\begin{figure*}[t!]
    \centering
     \begin{subfigure}[b]{1.0\textwidth}
         \centering
         \includegraphics[width=\textwidth]{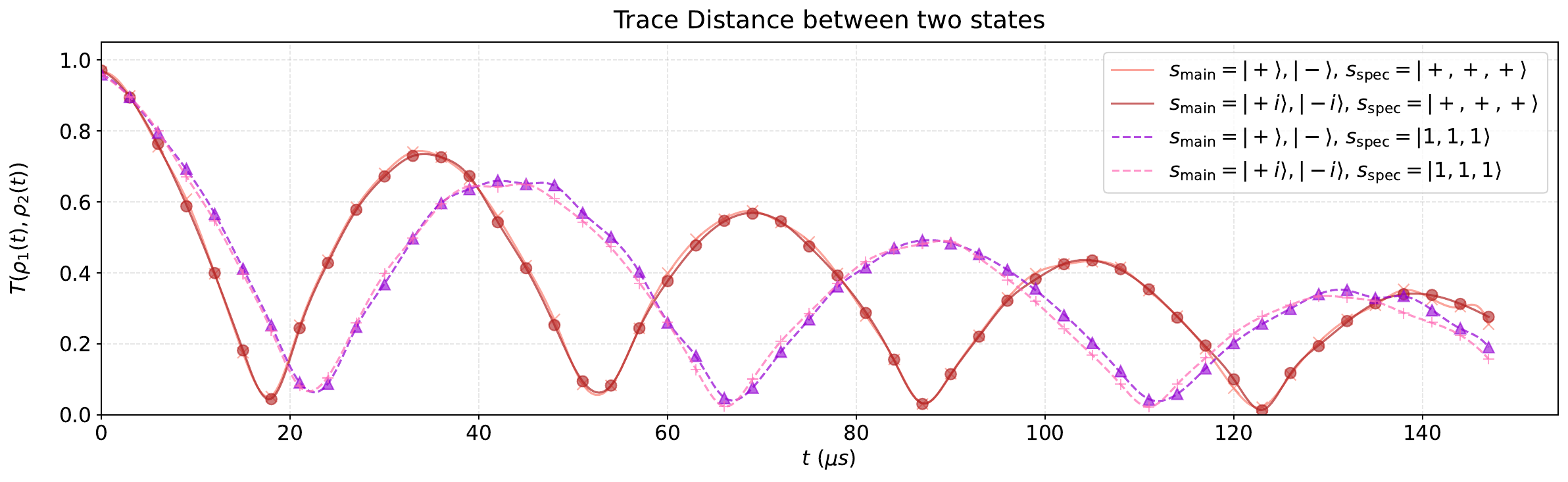}
     \end{subfigure}
     \begin{subfigure}[b]{1.0\textwidth}
         \centering
         \includegraphics[width=\textwidth]{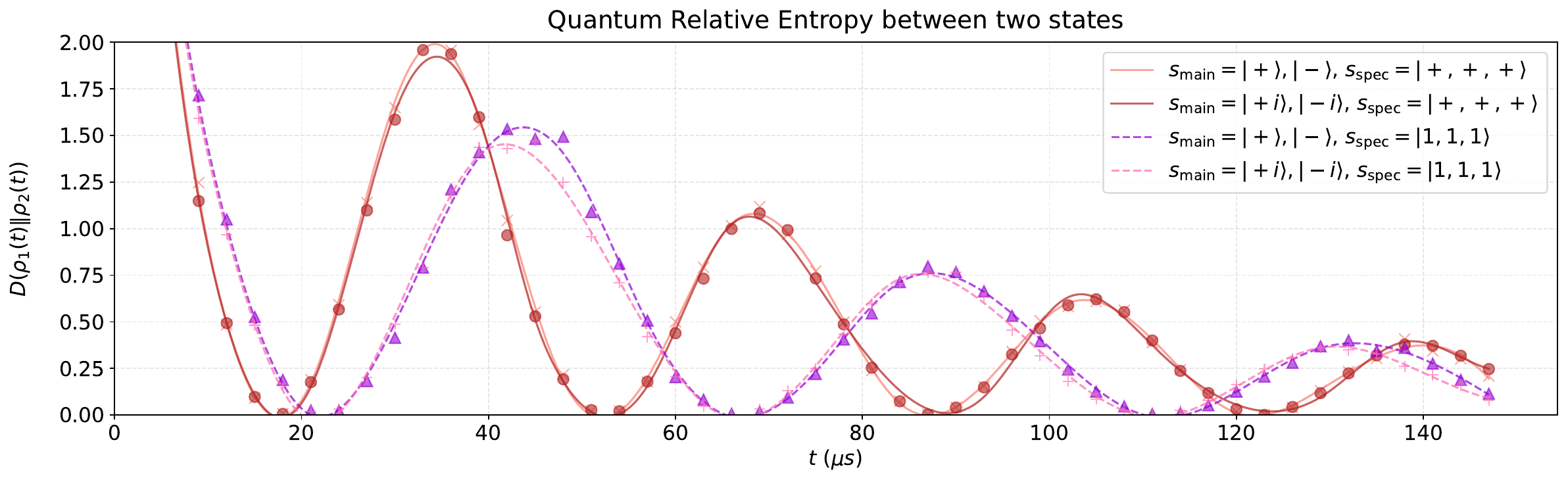}
     \end{subfigure}
     \caption{\justifying
     Time evolution of the trace-norm distance $T\bigl(\rho_1(t),\rho_2(t)\bigr)$ and the quantum relative entropy $D\bigl(\rho_1(t)\Vert\rho_2(t)\bigr)$ for the two initial main-qubit pairs $(\ket{+},\ket{-})$ and $(\ket{+i},\ket{-i})$. The experiments are conducted on IBM's 127-qubit ``Brussels'' processor, with main qubit 49 and spectator qubits $\{48,50,55\}$. Solid lines correspond to spectator qubits initialized in $\ket{+}^{\otimes3}$, and dashed lines correspond to spectator qubits initialized in $\ket{1}^{\otimes3}$. The periodic revivals indicate information backflow and hence non-Markovian behavior of the reduced main-qubit dynamics.}
  \label{fig:trace-dist-and-entropy}
\end{figure*}

\Cref{fig:trace-dist-and-entropy} shows the time evolution of the trace-norm distance and quantum relative entropy. Both quantities exhibit pronounced revivals, providing evidence for non-Markovian information backflow into the main qubit. At the same time, the revival amplitudes decay over time, consistent with the underlying relaxation dynamics that drive all trajectories toward the ground state $\ket{0}$, where the two evolved states become indistinguishable and both $T$ and $D$ vanish asymptotically.

\subsection{Information-theoretic probe of qubit crosstalk}
\label{sec:results-qmi}

A convenient way to quantify correlations generated between two idle qubits $A$ and $B$ is the quantum mutual information $0\le I(A:B)_{\rho(t)}\le2$, as discussed in \cref{sec:qmi-crosstalk}.
A positive value witnesses correlations between the two qubits, though not necessarily entanglement or a unique microscopic mechanism.

We performed four independent runs on IBM's 127-qubit ``Sherbrooke'' processor, each starting from one of the equatorial product states
\begin{align}
  \rho_{AB}(0) = \ketb{pp}{pp},\quad p\in\{+,-,+i,-i\} .
\end{align}
Both qubits are left idle for a variable duration $t$. At each time point, we perform full two-qubit state tomography, reconstruct $\rho_{AB}(t)$, and compute $I(A:B)_{\rho(t)}$. 

\begin{figure*}[ht!]
    \centering
    \begin{subfigure}[b]{\textwidth}
         \centering
         \includegraphics[width=1.0\textwidth]{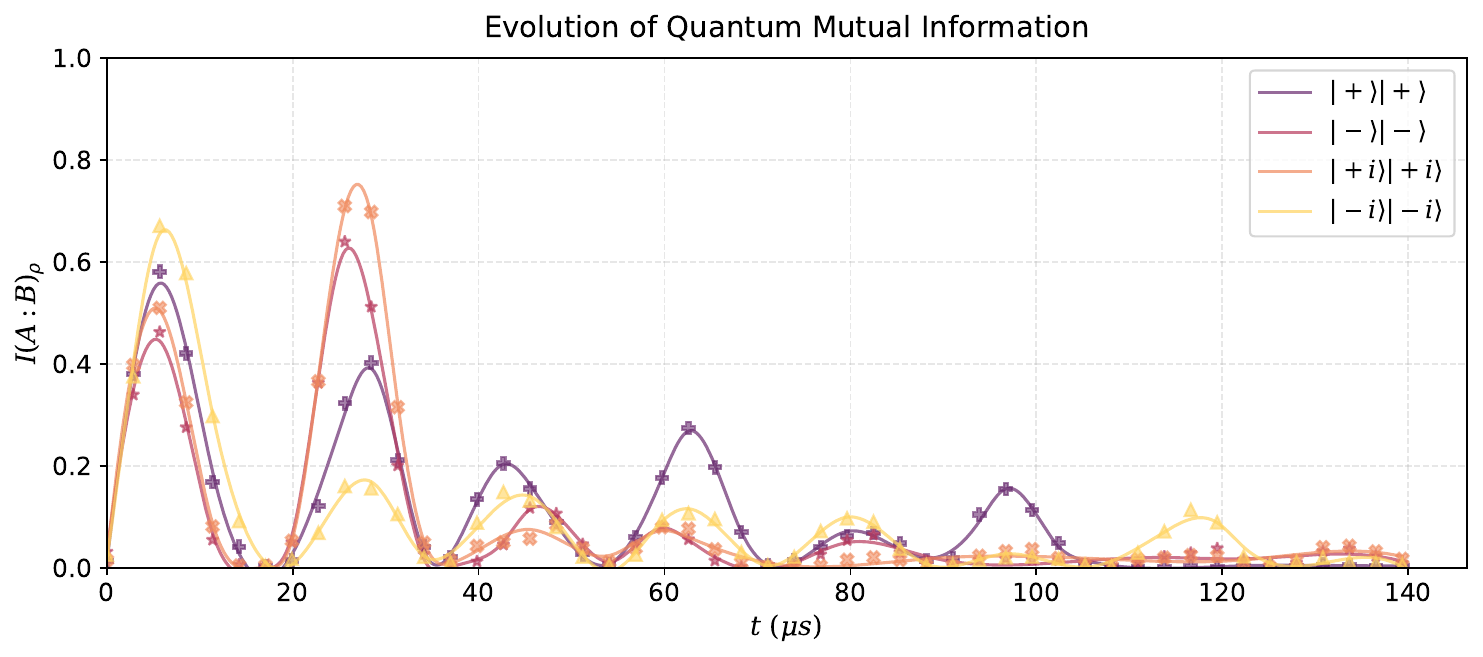}
     \end{subfigure}
    \caption{\justifying
    Time evolution of the quantum mutual information $I(A:B)_{\rho(t)}$ for four different initial product states.
    The rapid initial rise shows fast correlation buildup between the two qubits, while the subsequent revivals at approximately $10\mu\mathrm{s}$, $25\mu\mathrm{s}$, $45\mu\mathrm{s}$, and $65\mu\mathrm{s}$ indicate repeated generation of correlations during idle evolution.
    The decaying envelope is consistent with relaxation dynamics that suppress the generated correlations at long times.}
    \label{fig:qmi}
\end{figure*}

As shown in \cref{fig:qmi}, $I(A:B)_{\rho(t)}$ quickly departs from its initial baseline, reaches a first peak at short time, and exhibits pronounced revivals near $10\mu\mathrm{s}$, $25\mu\mathrm{s}$, $45\mu\mathrm{s}$, and $65\mu\mathrm{s}$. This behavior indicates repeated correlation buildup between nearby qubits and is consistent with a crosstalk-induced contribution to the observed memory effects. The overall envelope decays at long times, consistent with amplitude damping gradually driving the two-qubit state toward $\ket{0}^{\otimes2}$ and suppressing the generated correlations.

The four initial states in \cref{fig:qmi} exhibit different peak heights and revival patterns. These four equatorial states are related by local $Z$ rotations: writing
\beq
\ket{\psi_\phi}
=
\frac{\ket{0}+e^{i\phi}\ket{1}}{\sqrt2}
=
e^{i\phi/2}e^{-i\phi Z/2}\ket{+},
\eeq
we see that the states $\ket{+}$, $\ket{-}$, $\ket{+i}$, and $\ket{-i}$ correspond to different choices of $\phi$. Since local $Z$ rotations commute with an ideal $ZZ$ Hamiltonian, and since local amplitude damping and pure dephasing are phase-covariant, such a model would predict QMI curves that are identical up to local unitary rotations:
\beq
\begin{aligned}
I(A:B)_{\rho_{\phi_A,\phi_B}(t)}
=
I(A:B)_{\rho_{0,0}(t)} .
\end{aligned}
\eeq
Thus the distinct behavior of the four states indicates that the experiment is sensitive to additional nonidealities that break this simple phase covariance.

A natural explanation is that the effective two-qubit idle dynamics contains small terms beyond the diagonal $ZZ$ interaction, together with state-preparation, readout, and calibration imperfections. Residual transverse or mixed-axis couplings, such as $ZX$, $ZY$, $XZ$, $YZ$, $XX$, or $YY$ components, are not invariant under rotations of the initial equatorial axes, and could explain the observed differences among qubits initialized along $+X$, $-X$, $+Y$, or $-Y$. Local frame errors, ac-Stark shifts, pulse-dependent preparation errors, and leakage can further rotate the prepared axes relative to these residual couplings. As a result, the same underlying device crosstalk can generate different QMI peak heights and revival phases for the four nominally equivalent equatorial preparations.

The common revival times across the curves shown in \cref{fig:qmi} are nevertheless consistent with a shared coherent interaction timescale, while the preparation-dependent amplitudes and phases suggest that our minimal spectator-$ZZ$ model captures the dominant memory timescale but not all symmetry-breaking details of the two-qubit idle dynamics.

\begin{figure*}[t]
    \centering
    \includegraphics[width=1.0\linewidth]{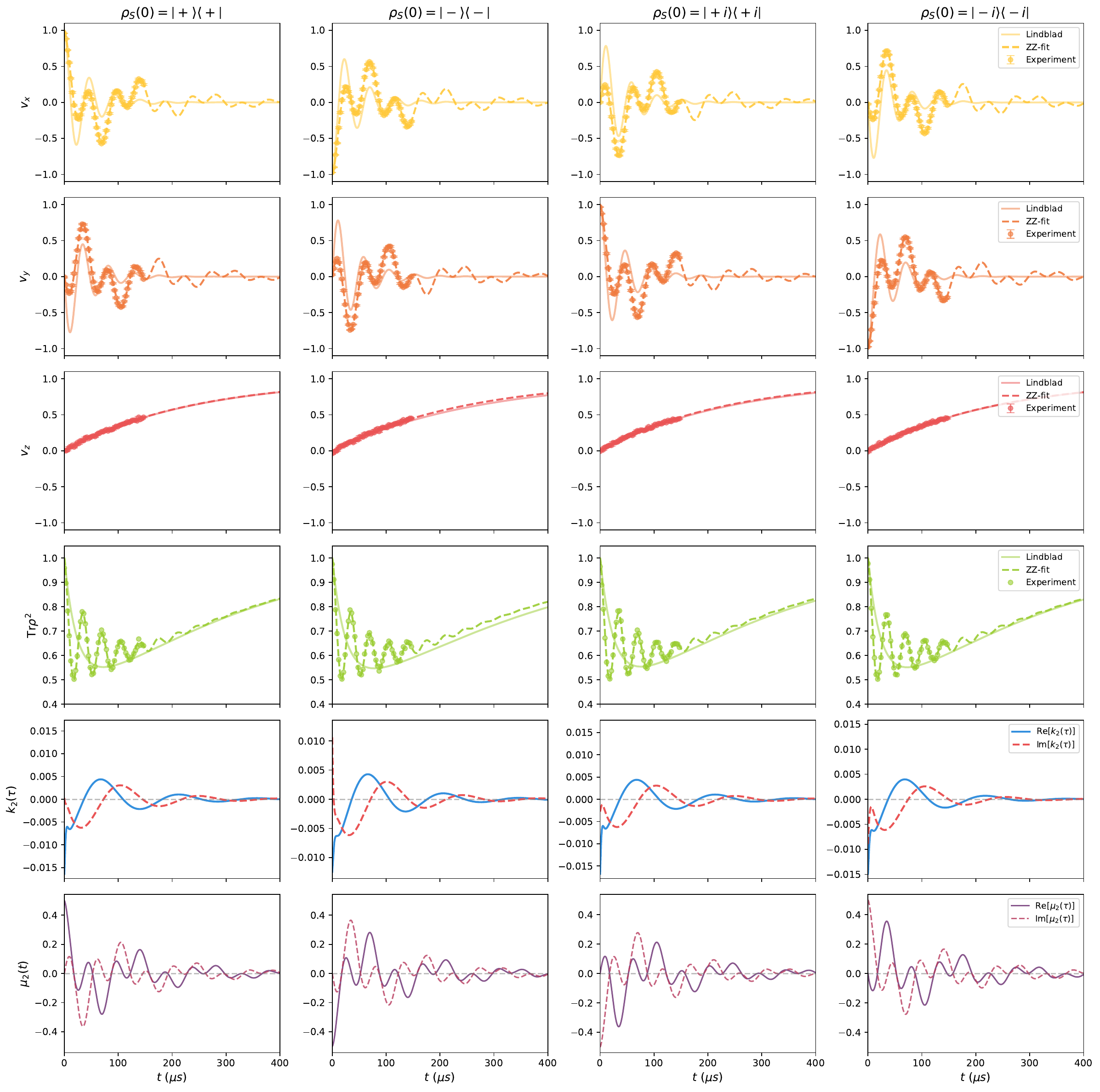}
    \caption{\justifying
    Reconstruction of the channel-resolved post-Markovian memory kernel for four different initial states of the main qubit. Each column corresponds to a different state preparation, indicated above the panels.
    \textbf{Rows 1-3} Measured Bloch-vector components $v_x(t)$, $v_y(t)$, and $v_z(t)$.
    Markers denote experimental data, solid lines denote the best-fit GKLS trajectories, and dashed lines denote the analytical spectator-$ZZ$ crosstalk fit. The GKLS model captures the overall relaxation behavior but does not reproduce the pronounced transverse revivals, whereas the microscopic $ZZ$ model tracks the observed oscillatory features.
    \textbf{Row 4} Purity $\Tr[\rho(t)^2]$ for the same data sets, again showing deviations from the purely Markovian prediction at short and intermediate times.
    \textbf{Row 5} Reconstructed transverse memory kernel $k_2(\tau)$ obtained from the analytical relation
    $\tilde{k}_2(z)=z-\lambda_2-1/\tilde{\xi}_2(z)$; the real and imaginary parts are shown separately.
    \textbf{Row 6} Corresponding transverse coefficient $\mu_2(t)$, whose oscillatory behavior reflects the same memory timescale encoded in the kernel.}
    \label{fig:pmme}
\end{figure*}

\subsection{Reconstruction of the PMME memory kernel}

To characterize the memory effects in the experimentally observed dynamics,
we reconstruct the channel-resolved memory kernel of the generalized
post-Markovian equation introduced in
\cref{eq:channel-memory-master-equation},
where $\mathcal{L}$ is the best-fit Lindblad generator that captures the
asymptotic Markovian baseline, while $\mathcal{K}(\tau)$ encodes the
transient non-Markovian correction.  
As shown in \cref{sec:PMME}, trace preservation fixes $k_0(t)=0$, and the microscopic spectator-$ZZ$ model predicts that the longitudinal mode remains Markovian, so that $k_1(t)=0$. Moreover, Hermiticity implies $k_3(t)=k_2(t)^*$, as derived in Appendix~\ref{app:complex-conjugate}. It therefore suffices to reconstruct the transverse memory kernel $k_2(t)$.

We proceed in three steps:
\begin{enumerate}[leftmargin=*]
    \item \textbf{Tomography and mode extraction.}
    For the four initial main-qubit preparations corresponding to
    $\ketb{+}{+}$, $\ketb{-}{-}$, $\ketb{+i}{+i}$, and $\ketb{-i}{-i}$, we perform state tomography at $N_t$ probe times. The idle duration is swept by varying the number of identity gates in the idle block; equivalently, the probe times are
    \beq
    t_n=n_{\mathrm{Id}}^{(n)}\tau_{\mathrm{Id}},
    \quad
    n=1,\ldots,N_t,
    \eeq
    where $\tau_{\mathrm{Id}}$ is the calibrated identity-gate duration and $n_{\mathrm{Id}}^{(n)}$ is the number of identity gates used for the $n$th probe time. At each probe time, we reconstruct the Bloch components $v_x(t)$, $v_y(t)$, and $v_z(t)$, together with the purity $\Tr[\rho(t)^2]$. From the reconstructed density matrix we extract the transverse coefficient
    \beq
    \mu_2(t)=\Tr[L_2\rho(t)]
    \eeq
    and, for preparations with $\mu_2(0)\ne0$, the normalized transverse mode function
    \beq
    \xi_2(t)=\frac{\mu_2(t)}{\mu_2(0)}.
    \eeq

    \item \textbf{Markovian baseline fit.}
    We fit a GKLS model to the measured Bloch trajectories by minimizing the mean-squared error
    \beq
      \mathrm{MSE}_{\mathrm L}
      =
      \frac{1}{3N_{\mathrm{prep}}N_t}
      \sum_{r=1}^{N_{\mathrm{prep}}}
      \sum_{i\in\{x,y,z\}}
      \sum_{n=1}^{N_t}
      \left| \Delta v_i^{(r)}(t_n) \right|^2
    \eeq
    where $r$ labels the initial preparation, $N_{\mathrm{prep}}=4$, 
\beq
      \Delta v_i^{(r)}(t_n) = v_i^{(r)}(t_n) - \hat v_{i,\mathrm L}^{(r)}(t_n),
\eeq
and $\hat v_{i,\mathrm L}^{(r)}(t_n)$ denotes the Bloch component generated by the fitted GKLS equation for the same initial state. This fit determines the baseline parameters entering $\mathcal{L}$ and, in particular, the transverse eigenvalue $\lambda_2$.

    \item \textbf{Analytical kernel reconstruction from the microscopic $ZZ$ model.}
    We next fit the experimentally extracted transverse dynamics $\xi_2(t)$ with the analytical spectator-$ZZ$ crosstalk model derived in \cref{sec:zz-crosstalk-model} and Appendix~\ref{app:ZZ-crosstalk}. Substituting the fitted closed-form expression for $\tilde{\xi}_2(z)$ into \cref{eq:kernel-s-domain} yields the Laplace-domain kernel
    \beq
    \tilde{k}_2(z)
    =
    z-\lambda_2-\frac{1}{\tilde{\xi}_2(z)}.
    \eeq
    Because $\tilde{\xi}_2(z)$ is known analytically within the fitted microscopic model, the inverse Laplace transform can be evaluated using the partial-fraction procedure described in Appendix~\ref{app:laplace}. This gives the time-domain kernel $k_2(t)=k_3(t)^*$.
\end{enumerate}

\Cref{fig:pmme} shows that the experimental deviations from the Markovian baseline are most pronounced in the two transverse modes, $v_x(t)$ and $v_y(t)$. In the first two rows, the measured transverse components exhibit clear oscillatory revivals, with prominent features at roughly $25\mu\mathrm{s}$, $55\mu\mathrm{s}$, $90\mu\mathrm{s}$, and $125\mu\mathrm{s}$, which are not captured by the best-fit GKLS model. By contrast, the analytical spectator-$ZZ$ crosstalk model closely tracks these revivals across all four initial preparations. In the third row, $v_z(t)$ remains nearly monotonic and is already well described by the Markovian baseline, consistent with the microscopic result that the longitudinal mode remains Markovian and hence $k_1(t)=0$. The fourth row shows that the purity also displays short- and intermediate-time oscillatory deviations from the GKLS prediction, indicating that the observed non-Markovian corrections are tied primarily to the transverse coherence dynamics rather than to population relaxation. The fifth row shows the reconstructed transverse kernels $k_2(\tau)$, whose real and imaginary parts exhibit damped oscillations. The sixth row shows the corresponding transverse coefficient $\mu_2(t)$, whose oscillatory behavior reflects the same memory timescale encoded in the kernel. Across the four initial preparations, the reconstructed kernels and transverse modes are similar, supporting the interpretation that the observed non-Markovianity is associated with device dynamics rather than being solely an artifact of a particular state preparation.

\section{Conclusions}
\label{sec:conclusion}

In this work, we presented an information-theoretic and experiment-driven characterization of non-Markovian noise in IBM superconducting qubits. By combining process tomography, distinguishability-based witnesses, two-qubit correlation probes, and a channel-resolved reconstruction of a PMME-inspired memory-kernel model, we obtained a consistent picture of how memory effects and spectator-induced crosstalk affect reduced single-qubit dynamics on present-day superconducting hardware. The CP-divisibility analysis revealed extended regions in the $(s,t)$ plane where the inferred intermediate Choi matrix has negative minimum eigenvalues, providing evidence for a breakdown of CP-divisibility in the reconstructed dynamics. 

The information-backflow diagnostics provide complementary evidence for non-Markovian behavior. For initially orthogonal transverse state pairs, both the trace distance and the quantum relative entropy exhibit pronounced revivals, indicating temporary increases in distinguishability that are incompatible with CP-divisible reduced dynamics. Two-qubit tomography further shows growth and revivals of quantum mutual information on related timescales. These observations are consistent with short-range qubit-qubit correlations and crosstalk being important contributors to the observed memory effects.

On the modeling side, we introduced a channel-resolved memory-kernel model in which distinct dynamical modes are assigned independent memory functions rather than being constrained by a single scalar kernel as in the original PMME~\cite{shabani_completely_2005}. Starting from a best-fit GKLS generator that captures the long-time Markovian baseline, we derived the reduced single-qubit dynamics generated by spectator-induced $ZZ$ coupling with local dissipation. The resulting closed-form model tracks the experimentally observed transverse Bloch-vector revivals while leaving the longitudinal relaxation mode Markovian, in agreement with the structure of the microscopic model. This analytical solution further enables reconstruction of the transverse memory kernel $k_2(t)$ within the fitted model. The reconstructed kernels are similar across the four initial preparations and exhibit damped oscillatory real and imaginary parts, providing a compact reduced description of the device memory beyond the vanishing-kernel correction associated with the fitted Markovian baseline.

Taken together, these results show that present-day superconducting devices can exhibit reduced dynamics that are not fully captured by standard $T_1/T_2$-based Markovian descriptions. More broadly, our work connects operational non-Markovianity diagnostics, microscopic spectator-crosstalk modeling, and reduced memory-kernel reconstruction within a single framework. This combination provides useful tools for diagnosing, modeling, and ultimately mitigating non-Markovian noise in near-term superconducting quantum hardware.

\acknowledgments
This material is based upon work supported by, or in part by, the U. S. Army Research Laboratory and the U.S. Army Research Office under contract/grant number W911NF2310255, and by the Office of Naval Research under Contract/Grant No. N0001-4-26-12092.

\bibliography{references.bib}

\clearpage

\appendix

\section{Relation between the superoperator matrix and the Choi matrix}
\label{app:superop-to-choi-matrix}

With the convention in \cref{eq:Choi},
\begin{align}
\chi_t
=
(\mathcal I\otimes \Phi_t)(\ketb{\gamma}{\gamma})
=
\sum_{ij}
\ketb{i}{j}\otimes
\Phi_t(\ketb{i}{j}).
\end{align}
For an arbitrary operator
\begin{align}
\rho=\sum_{ij}\rho_{ij}\ketb{i}{j},
\end{align}
we have
\beq
\label{eq:A3}
\begin{aligned}
\Tr_1\bigl[(\rho^T\otimes I)\chi_t\bigr]
&=
\sum_{ij}
\Tr\bigl(\rho^T\ketb{i}{j}\bigr)
\Phi_t(\ketb{i}{j})
\\
&=
\sum_{ij}
\rho_{ij}\Phi_t(\ketb{i}{j})
=
\Phi_t(\rho).
\end{aligned}
\eeq
Here the transpose is taken in the basis used to define $\ket{\gamma}$. Since the partial trace is cyclic with respect to operators acting only on the subsystem being traced out,
\begin{align}
\Tr_1\bigl[(\rho^T\otimes I)\chi_t\bigr]
=
\Tr_1\bigl[\chi_t(\rho^T\otimes I)\bigr].
\end{align}
Thus the inverse Choi-Jamio\l{}kowski relation for this convention is
\begin{align}
\Phi_t(\rho)
=
\Tr_1\bigl[(\rho^T\otimes I)\chi_t\bigr]
=
\Tr_1\bigl[\chi_t(\rho^T\otimes I)\bigr].
\end{align}

Let the matrix elements of $\chi_t$ be defined by
\begin{align}
(\chi_t)_{ij,kl}
=
(\bra{i}\otimes\bra{j})\chi_t(\ket{k}\otimes\ket{l}).
\end{align}
From
\begin{align}
\chi_t
=
\sum_{k\ell}
\ketb{k}{l}\otimes
\Phi_t(\ketb{k}{l}),
\end{align}
we obtain
\begin{align}
(\chi_t)_{ki,\ell j}
=
\bra{i}\Phi_t(\ketb{k}{l})\ket{j}.
\end{align}
Therefore, using \cref{eq:A3},
\begin{align}
\begin{aligned}
[\Phi_t(\rho)]_{ij}
&=
\sum_{k\ell}
\rho_{k\ell}
\bra{i}\Phi_t(\ketb{k}{l})\ket{j}
\\
&=
\sum_{k\ell}
(\chi_t)_{ki,\ell j}\rho_{k\ell}.
\end{aligned}
\end{align}

Now define the Liouville representation by
\begin{align}
\operatorname{vec}\bigl(\Phi_t(\rho)\bigr)
=
S_t\operatorname{vec}(\rho),
\end{align}
with column-stacking vectorization and component labels
\begin{align}
[\operatorname{vec}(\rho)]_{ij}=\rho_{ij}.
\end{align}
Then
\begin{align}
[\operatorname{vec}(\Phi_t(\rho))]_{ij}
=
\sum_{k\ell}
(S_t)_{ij,k\ell}
[\operatorname{vec}(\rho)]_{k\ell}.
\end{align}
Comparing this equation with the expression for $[\Phi_t(\rho)]_{ij}$ gives
\begin{align}
(S_t)_{ij,k\ell}
=
(\chi_t)_{ki,\ell j}.
\end{align}
Thus $S_t=\mathcal R(\chi_t)$, where $\mathcal R$ is the index-reshuffling map defined above. The inverse reshuffling is
\begin{align}
(\chi_t)_{ij,k\ell}
=
(S_t)_{j\ell,ik}.
\end{align}

For completeness, we also record the trace-preservation condition in this convention. Taking the trace of the inverse Choi-Jamio\l{}kowski relation gives
\begin{align}
\Tr[\Phi_t(\rho)]
=
\Tr\bigl[(\rho^T\otimes I)\chi_t\bigr]
=
\Tr\bigl[\rho^T\Tr_2(\chi_t)\bigr].
\end{align}
Thus $\Tr[\Phi_t(\rho)]=\Tr(\rho)$ for all $\rho$ if and only if
\begin{align}
\Tr_2\chi_t=I.
\end{align}

We now derive the intermediate-map formula. In Liouville representation,
\begin{align}
|\rho(t)\rangle\!\rangle
=
S_t|\rho(0)\rangle\!\rangle .
\end{align}
If the dynamics is CP-divisible, then for every $t\ge s$ there exists a CPTP intermediate map $\mathcal V_{t,s}$ satisfying
\begin{align}
\Phi_t=\mathcal V_{t,s}\circ\Phi_s .
\end{align}
Let $S_{t,s}$ denote the Liouville representation of $\mathcal V_{t,s}$. Since composition of maps is represented by matrix multiplication,
\begin{align}
S_t=S_{t,s}S_s .
\end{align}
Whenever $S_s$ is invertible,
\begin{align}
S_{t,s}=S_tS_s^{-1}.
\end{align}
Consequently, the Choi matrix of the intermediate map is
\begin{align}
\chi_{t,s}
=
\mathcal R^{-1}(S_{t,s})
=
\mathcal R^{-1}(S_tS_s^{-1}).
\end{align}

If $S_s$ is not invertible, the equation
\begin{align}
X S_s=S_t
\end{align}
may have no solution or multiple solutions. A linear solution exists only if the action encoded by $S_t$ is consistent on the kernel of $S_s$; equivalently,
\begin{align}
S_t=S_tS_s^+S_s,
\end{align}
where $S_s^+$ is the Moore-Penrose pseudoinverse. When solutions exist, the pseudoinverse choice
\begin{align}
X=S_tS_s^+
\end{align}
is only one possible linear extension. The CP-divisibility question is instead the feasibility problem of finding a Choi matrix $\chi_{t,s}$ such that
\begin{align}
\mathcal R(\chi_{t,s})S_s=S_t,
\quad
\chi_{t,s}\ge0,
\quad
\Tr_2\chi_{t,s}=I .
\end{align}
Thus, for singular or ill-conditioned $S_s$, the pseudoinverse construction should be regarded as a diagnostic, unless it is supplemented by conditioning and uncertainty checks, or by a direct feasibility test for a CPTP extension.

Finally, note that if the experimental coordinates use $s$ for the start time and $\tau$ for the elapsed idle time, then the final absolute time is $s+\tau$, and the corresponding intermediate Choi matrix is
\begin{align}
\chi_{s+\tau,s}
=
\mathcal R^{-1}(S_{s+\tau}S_s^{-1}).
\end{align}

\section{Complex-conjugation relation between $k_2(t)$ and $k_3(t)$}
\label{app:complex-conjugate}

Let the Laplace transform be defined by
$\tilde g(z)=\int_0^\infty dt e^{-zt}g(t)$,
with $\operatorname{Re}z$ chosen in the region of convergence. Then
\beq
\begin{aligned}
\operatorname{Lap}\bigl[g^*(t)\bigr](z)
&=
\int_0^\infty dt e^{-zt}g^*(t)
\\
&=
\left[
\int_0^\infty dt e^{-z^*t}g(t)
\right]^*
=
\bigl[\tilde g(z^*)\bigr]^* .
\label{eq:laplace-complex-conjugate}
\end{aligned}
\eeq

For the single-qubit damped basis used in \cref{sec:PMME}, the two transverse eigenvalues obey $\lambda_3=\lambda_2^*$.
Hermiticity of $\rho(t)$ implies, using \cref{eq:RLlambda,eq:rho-expand}
\beq
\rho(t)=\mu^*_0(t)R_0 + \mu^*_1(t)R_1 + \mu^*_2(t)R_3 + \mu^*_3(t)R_2,
\eeq
and hence that the transverse mode amplitudes satisfy
\begin{align}
\mu_3(t)=\mu_2(t)^* .
\end{align}
For initial states with nonzero transverse coherence, using \cref{eq:xi} this gives
\begin{align}
\xi_3(t)=\xi_2(t)^*,
\end{align}
and therefore, by \cref{eq:laplace-complex-conjugate},
\begin{align}
\tilde{\xi}_3(z)
=
\bigl[\tilde{\xi}_2(z^*)\bigr]^* .
\label{eq:xi2-xi3-conjugate}
\end{align}

Using \cref{eq:laplace-complex-conjugate}, the Laplace transform of $k_2(t)^*$ is
\beq
\begin{aligned}
\operatorname{Lap}\bigl[k_2(t)^*\bigr](z)
&=
\bigl[\tilde{k}_2(z^*)\bigr]^*
\\
&=
z-\lambda_2^*
-\frac{1}{\bigl[\tilde{\xi}_2(z^*)\bigr]^*}
\\
&=
z-\lambda_3-\frac{1}{\tilde{\xi}_3(z)}
=
\tilde{k}_3(z),
\end{aligned}
\eeq
where we used \cref{eq:kernel-s-domain,eq:xi2-xi3-conjugate}. By uniqueness of the inverse Laplace transform in the common region of convergence, we conclude that
\begin{align}
k_3(t)=k_2(t)^* ,
\end{align}
as claimed in \cref{sec:PMME}.

\section{Spectator-$ZZ$ crosstalk model}
\label{app:ZZ-crosstalk}

This appendix derives the closed-form reduced dynamics of a designated main qubit coupled to spectator qubits through residual $ZZ$ interactions. We proceed in three steps:
\begin{enumerate}[label=(\roman*),leftmargin=0.5cm]
    \item \textbf{Pure $ZZ$ coupling.}
    We first consider the minimal coherent model in which the main qubit interacts with a single spectator qubit through a $ZZ$ Hamiltonian, without decoherence.

    \item \textbf{$ZZ$ coupling with local dissipation.}
    We then include local amplitude damping and pure dephasing on the main and spectator qubits.

    \item \textbf{Multiple spectators with local dissipation.}
    Finally, we generalize to a main qubit coupled through $ZZ$ interactions to $N$ spectators, with local Markovian dissipation acting on every qubit.
\end{enumerate}
At each step, we solve the Heisenberg-picture equations of motion for a closed operator set and extract the resulting main-qubit Bloch-vector dynamics.

\subsection{Pure $ZZ$ crosstalk between two qubits}

The simplest model of $ZZ$ crosstalk is the closed-system evolution generated by
\begin{align}
    H=\frac{1}{2}JZ_0Z_1 .
\end{align}
To monitor how this coherent interaction affects the main qubit, we track its Bloch-vector operators in the Heisenberg picture. Operators evolve according to
\begin{align}
    \frac{d}{dt}O(t)
    =
    \mathcal L_H^\dagger[O(t)],
    \quad
    \mathcal L_H(\rho)=-i[H,\rho],
\end{align}
so that
\begin{align}
    \mathcal L_H^\dagger(O)=i[H,O].
\end{align}
Applying $\mathcal L_H^\dagger$ to the relevant Pauli operators gives
\begin{align}
    \mathcal L_H^\dagger(X_0)
    &=
    -JY_0Z_1,
    &
    \mathcal L_H^\dagger(Y_0Z_1)
    &=
    JX_0,
    \\
    \mathcal L_H^\dagger(Y_0)
    &=
    JX_0Z_1,
    &
    \mathcal L_H^\dagger(X_0Z_1)
    &=
    -JY_0 .
\end{align}
Thus the operator set
\begin{align}
    \mathcal O
    =
    \{X_0,Y_0Z_1,Y_0,X_0Z_1\}
\end{align}
is closed under the Hamiltonian dynamics. Writing
\begin{align}
    \mathbf v
    =
    \begin{pmatrix}
        X_0\\
        Y_0Z_1\\
        Y_0\\
        X_0Z_1
    \end{pmatrix},
\end{align}
the equations of motion take the form
\begin{align}
    \dot{\mathbf v}(t)=M\mathbf v(t),
\end{align}
with
\begin{align}
    M
    =
    \begin{pmatrix}
        0 & -J & 0 & 0 \\
        J & 0 & 0 & 0 \\
        0 & 0 & 0 & J \\
        0 & 0 & -J & 0
    \end{pmatrix}.
\end{align}
The formal solution is $\mathbf v(t)=e^{Mt}\mathbf v(0)$. Since $M$ splits into two independent $2\times2$ blocks, we obtain
\begin{align}
    \langle X_0\rangle(t)
    &=
    \langle X_0\rangle_0\cos(Jt)
    -
    \langle Y_0Z_1\rangle_0\sin(Jt),
    \\
    \langle Y_0\rangle(t)
    &=
    \langle Y_0\rangle_0\cos(Jt)
    +
    \langle X_0Z_1\rangle_0\sin(Jt).
\end{align}
Consequently, the main-qubit Bloch components $v_x(t)=\langle X_0\rangle(t)$ and $v_y(t)=\langle Y_0\rangle(t)$ acquire oscillatory contributions at angular frequency $J$. The oscillation amplitudes depend both on the initial main-qubit transverse components and on the initial main-spectator correlators.

\subsection{$ZZ$ crosstalk between two qubits with local dissipation}

We now supplement the coherent $ZZ$ interaction with detuning and local Markovian noise. 
The full two-qubit Lindbladian is
\begin{align}
\mathcal L
=
\mathcal L_H
+
\sum_{q=0}^1
\left(
\gamma_{\downarrow,q}\mathcal D_{\sigma_q^-}
+
\gamma_{\phi,q}\mathcal D_{Z_q}
\right),
\end{align}
where 
\beq
\label{eq:H2q}
H  = -\frac{1}{2}\omega_0Z_0 + \frac{1}{2}JZ_0Z_1 .
\eeq
Define
\begin{align}
    \Gamma_0
    =
    \frac{\gamma_{\downarrow,0}}{2}
    +
    2\gamma_{\phi,0},
    \quad
    \Gamma_1
    =
    \gamma_{\downarrow,1}.
\end{align}
Evaluating $\mathcal L^\dagger$ on the closed operator set
\begin{align}
\mathcal O=\{X_0,Y_0Z_1,Y_0,X_0Z_1\}
\end{align}
gives
\begin{align}
\begin{aligned}
  \mathcal L^\dagger(X_0)
  &=
  -\Gamma_0X_0
  -JY_0Z_1
  +\omega_0Y_0,
  \\
  \mathcal L^\dagger(Y_0Z_1)
  &=
  JX_0
  -(\Gamma_0+\Gamma_1)Y_0Z_1
  +\Gamma_1Y_0
  -\omega_0X_0Z_1,
  \\
  \mathcal L^\dagger(Y_0)
  &=
  -\omega_0X_0
  -\Gamma_0Y_0
  +JX_0Z_1,
  \\
  \mathcal L^\dagger(X_0Z_1)
  &=
  \Gamma_1X_0
  +\omega_0Y_0Z_1
  -JY_0
  -(\Gamma_0+\Gamma_1)X_0Z_1 .
\end{aligned}
\end{align}
Thus $\dot{\mathbf v}=M\mathbf v$, with
\begin{align}
  M
  =
  \begin{pmatrix}
    -\Gamma_0 & -J & \omega_0 & 0 \\
    J & -\Gamma_0-\Gamma_1 & \Gamma_1 & -\omega_0 \\
    -\omega_0 & 0 & -\Gamma_0 & J \\
    \Gamma_1 & \omega_0 & -J & -\Gamma_0-\Gamma_1
  \end{pmatrix}.
\end{align}
The spectator amplitude-damping term proportional to $\Gamma_1$ couples the two $2\times2$ blocks that were independent in the purely coherent case.

\paragraph{Block diagonalization.}
Although $M$ is no longer block diagonal in the Pauli-product basis above, it block diagonalizes in the transverse ladder basis
\begin{align}
  \Sigma_0^\pm
  =
  \frac{1}{2}(X_0\mp iY_0),
  \quad
  \Sigma_1^\pm
  =
  \frac{1}{2}(X_0Z_1\mp iY_0Z_1).
\end{align}
Let
\begin{align}
  \mathbf w
  =
  U_\Sigma\mathbf v,
  \quad
  \mathbf w
  =
  \begin{pmatrix}
  \Sigma_0^+\\
  \Sigma_1^+\\
  \Sigma_0^-\\
  \Sigma_1^-
  \end{pmatrix},
\end{align}
with
\begin{align}
  U_\Sigma
  =
  \frac{1}{2}
  \begin{pmatrix}
    1 & 0 & -i & 0 \\
    0 & -i & 0 & 1 \\
    1 & 0 & i & 0 \\
    0 & i & 0 & 1
  \end{pmatrix}.
\end{align}
Then
\begin{align}
  \tilde M
  =
  U_\Sigma M U_\Sigma^{-1}
  =
  \tilde M^+\oplus\tilde M^-,
\end{align}
where
\begin{align}
\tilde M^\pm
=
  \begin{pmatrix}
    -\Gamma_0\pm i\omega_0 & \mp iJ \\
    \Gamma_1\mp iJ & -\Gamma_0-\Gamma_1\pm i\omega_0
  \end{pmatrix}.
\end{align}
The $\tilde M^-$ block is the complex conjugate of the $\tilde M^+$ block. Therefore, the $\tilde M^+$ block determines the dynamics of the transverse coherence, with the $\tilde M^-$ dynamics following by complex conjugation.

\paragraph{Matrix exponential.}
To compute $e^{\tilde M^+t}$, decompose $\tilde M^+$ into traceful and traceless parts:
\begin{align}
  \tilde M^+
  &=
  \left(
  -\Gamma_0-\frac{\Gamma_1}{2}+i\omega_0
  \right)I
  +
  B^+,
  \\
  B^+
  &=
  \begin{pmatrix}
     \frac{\Gamma_1}{2} & -iJ \\
     \Gamma_1-iJ & -\frac{\Gamma_1}{2}
  \end{pmatrix}.
\end{align}
Let
\begin{align}
\Omega
=
\frac{\Gamma_1}{2}-iJ.
\end{align}
Then $\operatorname{Tr}(B^+)=0$, $\det B^+=-\Omega^2$, and
\begin{align}
(B^+)^2=\Omega^2I.
\end{align}
It follows that
\begin{align}
    e^{B^+t}
    =
    \cosh(\Omega t)I
    +
    \frac{\sinh(\Omega t)}{\Omega}B^+.
\end{align}
Thus
\begin{align}
    e^{\tilde M^+t}
    =
    e^{-\left(\Gamma_0+\frac{\Gamma_1}{2}-i\omega_0\right)t}
    \begin{pmatrix}
        C_\Omega+\frac{\Gamma_1}{2\Omega}S_\Omega
        &
        -i\frac{J}{\Omega}S_\Omega
        \\[0.25cm]
        \frac{\Gamma_1-iJ}{\Omega}S_\Omega
        &
        C_\Omega-\frac{\Gamma_1}{2\Omega}S_\Omega
    \end{pmatrix},
\end{align}
where
\begin{align}
C_\Omega=\cosh(\Omega t),
\quad
S_\Omega=\sinh(\Omega t).
\end{align}

\paragraph{Initial spectator in $\ket{+}$.}
Suppose the main qubit is prepared in $\ket{+}$ and the spectator qubit is also prepared in $\ket{+}$. Then
\begin{align}
    \langle \Sigma_0^+\rangle_0
    =
    \frac{1}{2},
    \quad
    \langle \Sigma_1^+\rangle_0
    =
    0,
\end{align}
so that
\begin{align}
    \mathbf w_+(0)
    =
    \begin{pmatrix}
    \frac{1}{2}\\
    0
    \end{pmatrix}.
\end{align}
Therefore,
\begin{align}
    \langle \Sigma_0^+\rangle(t)
    =
    \frac{1}{2}
    e^{-\left(\Gamma_0+\frac{\Gamma_1}{2}-i\omega_0\right)t}
    \left[
    \cosh(\Omega t)
    +
    \frac{\Gamma_1}{2\Omega}\sinh(\Omega t)
    \right].
\end{align}
The main-qubit Bloch components are
\begin{align}
    v_x(t)
    =
    2\operatorname{Re}\langle\Sigma_0^+\rangle(t),
    \quad
    v_y(t)
    =
    -2\operatorname{Im}\langle\Sigma_0^+\rangle(t).
\end{align}

In the weak-spectator-damping limit $\Gamma_1/(2J)\ll1$, one has $\Omega\approx -iJ$, and the leading behavior is
\begin{align}
    v_x(t)
    &\approx
    e^{-\left(\Gamma_0+\frac{\Gamma_1}{2}\right)t}
    \cos(\omega_0t)\cos(Jt),
    \\
    v_y(t)
    &\approx
    -e^{-\left(\Gamma_0+\frac{\Gamma_1}{2}\right)t}
    \sin(\omega_0t)\cos(Jt).
\end{align}
Thus the $ZZ$ crosstalk term induces an additional oscillatory modulation at frequency $J$.

In the opposite limit $\Gamma_1/(2J)\gg1$, the spectator relaxes rapidly to the ground state. After a transient of order $\Gamma_1^{-1}$, the dominant contribution is
\begin{align}
    v_x(t)
    &\approx
    e^{-\Gamma_0t}
    \cos\left[(\omega_0-J)t\right],
    \\
    v_y(t)
    &\approx
    -e^{-\Gamma_0t}
    \sin\left[(\omega_0-J)t\right],
\end{align}
up to corrections suppressed by $J/\Gamma_1$ and by the fast transient $e^{-\Gamma_1t}$.

\paragraph{Initial spectator in $\ket{0}$.}
Now suppose the main qubit is prepared in $\ket{+}$ and the spectator qubit is prepared in $\ket{0}$. Then
\begin{align}
    \langle \Sigma_0^+\rangle_0
    =
    \frac{1}{2},
    \quad
    \langle \Sigma_1^+\rangle_0
    =
    \frac{1}{2},
\end{align}
so that
\begin{align}
    \mathbf w_+(0)
    =
    \begin{pmatrix}
    \frac{1}{2}\\
    \frac{1}{2}
    \end{pmatrix}.
\end{align}
Using the expression for $e^{\tilde M^+t}$,
\begin{align}
  \langle\Sigma_0^+\rangle(t)
  &=
  \frac{1}{2}
  e^{-\left(\Gamma_0+\frac{\Gamma_1}{2}\right)t}
  e^{i\omega_0t}
  \left[
  C_\Omega
  +
  \frac{\Gamma_1}{2\Omega}S_\Omega
  -
  i\frac{J}{\Omega}S_\Omega
  \right]
  \\
  &=
  \frac{1}{2}
  e^{-\left(\Gamma_0+\frac{\Gamma_1}{2}\right)t}
  e^{i\omega_0t}
  \left(C_\Omega+S_\Omega\right)
  \\
  &=
  \frac{1}{2}
  e^{-\Gamma_0t}
  e^{i(\omega_0-J)t},
\end{align}
where we used $C_\Omega+S_\Omega=e^{\Omega t}$. Hence
\begin{align}
  v_x(t)
  &=
  e^{-\Gamma_0t}
  \cos\left[(\omega_0-J)t\right],
  \\
  v_y(t)
  &=
  -e^{-\Gamma_0t}
  \sin\left[(\omega_0-J)t\right].
\end{align}
The spectator damping rate $\Gamma_1$ does not appear in the decay envelope in this case because the spectator qubit is already in the ground state.

\subsection{$ZZ$ crosstalk with $N$ spectator qubits}

We now generalize the single-spectator calculation to $N$ spectators. Consider
\begin{align}
    H
    =
    -\frac{1}{2}\omega_0Z_0
    +
    \sum_{q=1}^{N}\frac{1}{2}J_{0q}Z_0Z_q,
\end{align}
with local dissipators
\begin{align}
\mathcal L_D = \sum_{q=0}^{N}
\left(
\gamma_{\downarrow,q}\mathcal D_{\sigma_q^-}
+
\gamma_{\phi,q}\mathcal D_{Z_q}
\right).
\end{align}

\paragraph{Closed operator sets.}
The relevant closed operator sets are
\begin{align}
    \mathcal O_\pm
    =
    \left\{
    \Sigma_{\mathbf s}^{\pm}
    =
    \frac{1}{2}(X_0\mp iY_0)
    Z_1^{s_1}Z_2^{s_2}\cdots Z_N^{s_N}
    :
    \mathbf s\in\{0,1\}^N
    \right\}.
\end{align}
Here $\mathbf s=(s_1,\ldots,s_N)$ and $\mathbf e_q$ is the bit string with a $1$ in position $q$ and zeros elsewhere. The Hamiltonian contributions are
\begin{align}
    i\left[-\frac{1}{2}\omega_0Z_0,\Sigma_{\mathbf s}^{\pm}\right]
    &=
    \pm i\omega_0\Sigma_{\mathbf s}^{\pm},
    \\
    i\left[\frac{1}{2}J_{0q}Z_0Z_q,\Sigma_{\mathbf s}^{\pm}\right]
    &=
    \mp iJ_{0q}\Sigma_{\mathbf s\oplus\mathbf e_q}^{\pm}.
\end{align}
The amplitude-damping adjoint acts as
\begin{align}
    \mathcal D^\dagger_{\sigma_q^-}(\Sigma_{\mathbf s}^\pm)
    =
    \begin{cases}
      -\frac{1}{2}\Sigma_{\mathbf s}^\pm, & q=0,\\[0.2cm]
      \delta_{s_q,1}
      \left(
      \Sigma_{\mathbf s\oplus\mathbf e_q}^{\pm}
      -
      \Sigma_{\mathbf s}^{\pm}
      \right), & q\ge1,
    \end{cases}
\end{align}
and the pure-dephasing adjoint acts as
\begin{align}
    \mathcal D^\dagger_{Z_q}(\Sigma_{\mathbf s}^{\pm})
    =
    \begin{cases}
      -2\Sigma_{\mathbf s}^{\pm}, & q=0,\\[0.2cm]
      0, & q\ge1.
    \end{cases}
\end{align}
Thus each of $\mathcal O_+$ and $\mathcal O_-$ is closed under the full Lindbladian.

\paragraph{Parity-space representation.}
Order the components by the bit string $\mathbf s$ and identify
\begin{align}
    \Sigma_{\mathbf s}^+
    \longleftrightarrow
    \ket{\mathbf s}
    \in
    \mathscr H_+,
    \quad
    \Sigma_{\mathbf s}^-
    \longleftrightarrow
    \ket{\mathbf s}
    \in
    \mathscr H_-,
\end{align}
where
\begin{align}
    \ket{\mathbf s}
    =
    \ket{s_1}\otimes\cdots\otimes\ket{s_N}.
\end{align}
The spaces $\mathscr H_\pm$ are abstract parity spaces isomorphic to $(\mathbb C^2)^{\otimes N}$.\footnote{The $\mathscr H_\pm$ are not additional physical Hilbert spaces; they only encode the closed operator basis.}

In this representation, the generator on each transverse sector is a Kronecker sum:
\begin{align}
    \tilde M^\pm
    &=
    -\left(
    \Gamma_0
    +
    \frac{1}{2}\sum_{q=1}^{N}\Gamma_q
    \mp i\omega_0
    \right)I
    \nonumber\\
    &\hspace{1cm}
    +
    \sum_{q=1}^{N}
    I^{\otimes(q-1)}
    \otimes
    B_q^\pm
    \otimes
    I^{\otimes(N-q)},
\end{align}
where
\begin{align}
\Gamma_0
=
2\gamma_{\phi,0}
+
\frac{\gamma_{\downarrow,0}}{2},
\quad
\Gamma_q
=
\gamma_{\downarrow,q}
\quad
(q\ge1),
\end{align}
and the single-spectator block acting on the $q$th parity bit is
\begin{align}
    B_q^\pm
    =
    \begin{pmatrix}
        \frac{\Gamma_q}{2} & \mp iJ_{0q} \\
        \Gamma_q\mp iJ_{0q} & -\frac{\Gamma_q}{2}
    \end{pmatrix}.
\end{align}
This is the same $2\times2$ block obtained in the single-spectator calculation, with $J$ and $\Gamma_1$ replaced by $J_{0q}$ and $\Gamma_q$.

Since all terms in the Kronecker sum act on different parity bits, the propagator factorizes:
\begin{align}
    e^{\tilde M^\pm t}
    =
    e^{(-\Gamma_0\pm i\omega_0)t}
    \bigotimes_{q=1}^{N}
    \left(
    e^{-\frac{\Gamma_qt}{2}}
    e^{B_q^\pm t}
    \right).
\end{align}
Although the closed operator set has dimension $2^N$ in each transverse sector, the propagator reduces to a tensor product of single-spectator factors.

\paragraph{Spectators initialized in equatorial states.}
Suppose the main qubit is prepared in $\ket{+}$ and each spectator has zero initial $Z$ polarization, as for the states $\ket{+}$, $\ket{-}$, $\ket{+i}$, and $\ket{-i}$. For definiteness, take the spectator state to be $\ket{+}^{\otimes N}$. Then
\begin{align}
    \langle \Sigma_{\mathbf 0}^+\rangle_0
    =
    \frac{1}{2},
    \quad
    \langle \Sigma_{\mathbf s}^+\rangle_0
    =
    0
    \quad
    \text{for }
    \mathbf s\ne\mathbf 0,
\end{align}
so that
\begin{align}
    \mathbf w_+(0)
    =
    \frac{1}{2}\ket{\mathbf 0}.
\end{align}
The first component of $\mathbf w_+(t)$ is
\beq
\begin{aligned}
    \langle \Sigma_{\mathbf 0}^+\rangle(t)
    &=
    \left[
    e^{\tilde M^+t}\mathbf w_+(0)
    \right]_{\mathbf 0}
    \\
    &=
    \frac{1}{2}
    e^{(-\Gamma_0+i\omega_0)t}
    \prod_{q=1}^{N}
    e^{-\frac{\Gamma_qt}{2}}
    \bra{0}e^{B_q^+t}\ket{0}.
\end{aligned}
\eeq
For
\begin{align}
\Omega_q
=
\frac{\Gamma_q}{2}
-
iJ_{0q},
\quad
C_q=\cosh(\Omega_qt),
\quad
S_q=\sinh(\Omega_qt),
\end{align}
the required matrix element is
\begin{align}
    \bra{0}e^{B_q^+t}\ket{0}
    =
    C_q
    +
    \frac{\Gamma_q}{2\Omega_q}S_q.
\end{align}

In the weak-spectator-damping limit $\Gamma_q/(2J_{0q})\ll1$ for all $q$,
\begin{align}
    \bra{0}e^{B_q^+t}\ket{0}
    \approx
    \cos(J_{0q}t),
\end{align}
and hence
\beq
\begin{aligned}
    v_x(t)
    &\approx
    e^{-\Gamma_0t}
    \cos(\omega_0t)
    \prod_{q=1}^{N}
    e^{-\frac{\Gamma_qt}{2}}
    \cos(J_{0q}t),
    \\
    v_y(t)
    &\approx
    -e^{-\Gamma_0t}
    \sin(\omega_0t)
    \prod_{q=1}^{N}
    e^{-\frac{\Gamma_qt}{2}}
    \cos(J_{0q}t).
\end{aligned}
\eeq
In the opposite limit $\Gamma_q/(2J_{0q})\gg1$, each spectator rapidly relaxes to the ground state. After the fast transients have decayed, the dominant contribution is
\beq
\begin{aligned}
    v_x(t)
    &\approx
    e^{-\Gamma_0t}
    \cos\left(
    \omega_0t
    -
    \sum_{q=1}^{N}J_{0q}t
    \right),
    \\
    v_y(t)
    &\approx
    -e^{-\Gamma_0t}
    \sin\left(
    \omega_0t
    -
    \sum_{q=1}^{N}J_{0q}t
    \right),
\end{aligned}
\eeq
up to corrections suppressed by $J_{0q}/\Gamma_q$ and by fast transients $e^{-\Gamma_qt}$.

\paragraph{Spectators initialized in $\ket{0}^{\otimes N}$.}
Now suppose each spectator qubit is initialized in $\ket{0}$. Then
\begin{align}
    \langle \Sigma_{\mathbf s}^+\rangle_0
    =
    \frac{1}{2}
    \quad
    \text{for all }
    \mathbf s,
\end{align}
so that, in the parity space,
\begin{align}
    \mathbf w_+(0)
    =
    \frac{1}{2}
    \sum_{\mathbf s\in\{0,1\}^N}
    \ket{\mathbf s}
    =
    2^{\frac{N}{2}-1}\ket{+}^{\otimes N}.
\end{align}
The first component is
\begin{align}
    \langle \Sigma_{\mathbf 0}^+\rangle(t)
    =
    \frac{1}{2}
    e^{(-\Gamma_0+i\omega_0)t}
    \prod_{q=1}^{N}
    \left(
    \sqrt{2}
    e^{-\frac{\Gamma_qt}{2}}
    \bra{0}e^{B_q^+t}\ket{+}
    \right).
\end{align}
Using
\begin{align}
    \sqrt{2}\bra{0}e^{B_q^+t}\ket{+}
    =
    e^{\Omega_qt}
    =
    e^{\frac{\Gamma_qt}{2}}e^{-iJ_{0q}t},
\end{align}
we find
\beq
\begin{aligned}
    v_x(t)
    &=
    e^{-\Gamma_0t}
    \cos\left(
    \omega_0t
    -
    \sum_{q=1}^{N}J_{0q}t
    \right),
    \\
    v_y(t)
    &=
    -e^{-\Gamma_0t}
    \sin\left(
    \omega_0t
    -
    \sum_{q=1}^{N}J_{0q}t
    \right).
\end{aligned}
\eeq

\paragraph{General pure product initial states.}
Finally, consider arbitrary pure product initial states
\begin{align}
    \ket{\psi_q}
    =
    \cos\left(\frac{\theta_q}{2}\right)\ket{0}
    +
    e^{i\phi_q}
    \sin\left(\frac{\theta_q}{2}\right)\ket{1},
    \quad
    q=0,\ldots,N,
\end{align}
with
\begin{align}
    \ket{\Psi(0)}
    =
    \bigotimes_{q=0}^{N}\ket{\psi_q}.
\end{align}
Let
\begin{align}
x_q=\sin\theta_q\cos\phi_q,
\quad
y_q=\sin\theta_q\sin\phi_q,
\quad
z_q=\cos\theta_q
\end{align}
denote the initial Bloch-vector components of qubit $q$. Since the initial state factorizes,
\beq
\begin{aligned}
    \langle \Sigma_{\mathbf s}^{+}\rangle_0
    &=
    \frac{1}{2}
    \langle\psi_0|(X_0-iY_0)|\psi_0\rangle
    \prod_{q=1}^{N}
    \langle\psi_q|Z_q^{s_q}|\psi_q\rangle
    \\
    &=
    \frac{x_0-iy_0}{2}
    \prod_{q=1}^{N}
    z_q^{s_q}.
\end{aligned}
\eeq
Thus, in the parity space,
\begin{align}
    \mathbf w_+(0)
    =
    \frac{x_0-iy_0}{2}
    \bigotimes_{q=1}^{N}
    \left(\ket{0}+z_q\ket{1}\right).
\end{align}
Therefore,
\beq
\begin{aligned}
    \langle \Sigma_{\mathbf 0}^+\rangle(t)
    &=
    e^{(-\Gamma_0+i\omega_0)t}
    \frac{x_0-iy_0}{2}\times\\
    &\quad \prod_{q=1}^{N}
    e^{-\frac{\Gamma_qt}{2}}
    \bra{0}e^{B_q^+t}
    \left(\ket{0}+z_q\ket{1}\right)
    \\
    &=
    e^{(-\Gamma_0+i\omega_0)t}
    \frac{x_0-iy_0}{2}\times\\
    &\quad
    \prod_{q=1}^{N}
    e^{-\frac{\Gamma_qt}{2}}
    \left[
    C_q
    +
    \frac{\frac{\Gamma_q}{2}-iz_qJ_{0q}}{\Omega_q}S_q
    \right].
\end{aligned}
\eeq
The corresponding Bloch components are
\begin{align}
    v_x(t)
    =
    2\operatorname{Re}\langle\Sigma_{\mathbf 0}^+\rangle(t),
    \quad
    v_y(t)
    =
    -2\operatorname{Im}\langle\Sigma_{\mathbf 0}^+\rangle(t).
\end{align}
For an initial main-qubit state with nonzero transverse coherence, the normalized transverse mode function is therefore
\begin{align}
    \xi_2(t)
    =
    e^{(-\Gamma_0+i\omega_0)t}
    \prod_{q=1}^{N}
    e^{-\frac{\Gamma_qt}{2}}
    \left[
    C_q
    +
    \frac{\frac{\Gamma_q}{2}-iz_qJ_{0q}}{\Omega_q}S_q
    \right].
\end{align}
For the equatorial spectator preparations used in the main text, $z_q=0$, which reduces this expression to the factorized modulation used in \cref{sec:zz-crosstalk-model}, i.e., \cref{eq:xi2-1,eq:xi2-2}.

\section{Analytical Laplace inversion and initial-value constraints}
\label{app:laplace}

In this section, we describe the analytical inversion procedure used to reconstruct the transverse memory kernel $k_2(t)$ from the fitted spectator-$ZZ$ model. We first explain why the apparent polynomial terms in the Laplace-domain reconstruction formula cancel when the baseline generator captures the instantaneous Markovian contribution. We then apply the procedure to the three-spectator model used in the main text.

\subsection{Cancellation of polynomial terms from the initial-time behavior}

Assume that $\xi_i(t)$ is twice differentiable near $t=0$ and of exponential order, so that the following successive integrations by parts are valid for large $\operatorname{Re}z$:
\beq
\begin{aligned}
\tilde{\xi}_i(z)
&=
\int_0^\infty e^{-zt}\xi_i(t)dt
\\
&=
\frac{\xi_i(0)}{z}
+
\frac{1}{z}\int_0^\infty e^{-zt}\dot{\xi}_i(t)dt
\\
&=
\frac{\xi_i(0)}{z}
+
\frac{\dot{\xi}_i(0)}{z^2}
+
\frac{1}{z^2}\int_0^\infty e^{-zt}\ddot{\xi}_i(t)dt .
\end{aligned}
\eeq
If $\ddot{\xi}_i(t)$ is locally bounded near the origin and of exponential order, the last integral is $\mathcal O(1/z)$ as $z\to\infty$ in the right half-plane. Hence
\beq
\tilde{\xi}_i(z)
=
\frac{1}{z}
+
\frac{a_i}{z^2}
+
\mathcal O\left(\frac{1}{z^3}\right),
\eeq
where $a_i\coloneqq \dot{\xi}_i(0)$, and we used $\xi_i(0)=1$ [\cref{eq:xi}].

Taking the reciprocal yields
\beq
\label{eq:recip}
\begin{aligned}
\frac{1}{\tilde{\xi}_i(z)}
&=
z
\left(
1-\frac{a_i}{z}
+
\mathcal O\left(\frac{1}{z^2}\right)
\right)
\\
&=
z-a_i+\mathcal O\left(\frac{1}{z}\right).
\end{aligned}
\eeq
Substituting into \cref{eq:kernel-s-domain} gives
\beq
\begin{aligned}
\tilde{k}_i(z)
&=
z-\lambda_i
-
\left[
z-a_i+\mathcal O\left(\frac{1}{z}\right)
\right]
\\
&=
a_i-\lambda_i
+
\mathcal O\left(\frac{1}{z}\right).
\end{aligned}
\eeq
Thus $\tilde{k}_i(z)$ has no polynomial part if and only if
$a_i=\lambda_i$.
This equality holds for the memory-kernel equation when the kernel is locally integrable at the origin and all singular Markovian contributions have been included in the baseline generator $\mathcal L$. Indeed, in that case the convolution term vanishes at $t=0$, so that
\beq
\label{eq:cancel-cond}
\dot{\xi}_i(0)=\lambda_i.
\eeq
Under this assumption,
\beq
\tilde{k}_i(z)=\mathcal O\left(\frac{1}{z}\right),
\eeq
and the inverse Laplace transform contains no distributional contribution supported at $t=0$. For rational fitted forms of $\tilde{\xi}_i(z)$, the remaining strictly proper rational function can then be inverted by partial fractions.

If instead $a_i\ne\lambda_i$, then the constant term $a_i-\lambda_i$ in $\tilde{k}_i(z)$ corresponds to a singular contribution proportional to $\delta(t)$ in the time-domain kernel. Such a term indicates either that the chosen baseline generator does not capture the instantaneous derivative of the fitted mode function, or that the effective kernel contains a singular Markovian component that should be absorbed into $\mathcal L$. We give a detailed explanation of these considerations in \cref{app:sec-detailed}.

\subsection{Application to the three-spectator model}

We now apply this framework to the $N=3$ spectator model. For the equatorial spectator preparations used in the transverse-kernel reconstruction, 
the reduced transverse mode has the form [\cref{eq:xi2-1}]
\beq
\xi_2(t)
=
e^{\lambda_2t}\Phi(t),
\quad
\Phi(t)
=
\prod_{q=1}^3\Phi_q(t),
\eeq
where $\Phi_q(t)$ is given in \cref{eq:xi2-2}.
Each spectator modulation satisfies
\beq
\Phi_q(0)=1,
\quad
\dot{\Phi}_q(0)=0.
\eeq
Consequently,
\beq
\label{eq:init-cond-Phi}
\Phi(0)=1,
\quad
\dot{\Phi}(0)=0,
\eeq
and hence
\beq
\dot{\xi}_2(0)=\lambda_2.
\eeq
\Cref{eq:cancel-cond} is therefore satisfied.

Expanding each spectator modulation into exponentials gives
\beq
\Phi_q(t)
=
c_{q,+}e^{\nu_{q,+}t}
+
c_{q,-}e^{\nu_{q,-}t},
\eeq
with
\beq
\begin{aligned}
c_{q,+}
&=
\frac{1}{2}
\left(
1+\frac{\Gamma_q}{2\Omega_q}
\right),
&
c_{q,-}
&=
\frac{1}{2}
\left(
1-\frac{\Gamma_q}{2\Omega_q}
\right),
\\
\nu_{q,+}
&=
-\frac{\Gamma_q}{2}+\Omega_q,
&
\nu_{q,-}
&=
-\frac{\Gamma_q}{2}-\Omega_q.
\end{aligned}
\eeq
Multiplying the three factors gives a sum of $2^3=8$ exponential terms,
\beq
\label{eq:Phi-sum}
\Phi(t)
=
\sum_{m=1}^8 A_m e^{\Lambda_m t}.
\eeq
Each $\Lambda_m$ is a sum
\beq
\Lambda_m
=
\nu_{1,\eta_1}
+
\nu_{2,\eta_2}
+
\nu_{3,\eta_3},
\quad
\eta_q\in\{+,-\},
\eeq
and $A_m$ is the corresponding product
\beq
A_m
=
c_{1,\eta_1}c_{2,\eta_2}c_{3,\eta_3}.
\eeq
The initial conditions in \cref{eq:init-cond-Phi} imply
\beq
\label{eq:init-cond-Phi-2}
\sum_{m=1}^8 A_m=1,
\quad
\sum_{m=1}^8 A_m\Lambda_m=0.
\eeq

Introducing the shifted Laplace variable $x=z-\lambda_2$, and using \cref{eq:Phi-sum}, we have
\beq
\begin{aligned}
\tilde{\xi}_2(z)
&=
\int_0^\infty e^{-zt}e^{\lambda_2t}\Phi(t)dt=
\int_0^\infty e^{-xt}\Phi(t)dt
\\
&=
\sum_{m=1}^8\frac{A_m}{x-\Lambda_m}
\equiv
\frac{P(x)}{Q(x)}.
\end{aligned}
\eeq
Before any possible cancellation of common factors,
\beq
Q(x)
=
\prod_{m=1}^8(x-\Lambda_m)
\eeq
is a monic polynomial of degree $8$, and
\beq
P(x)
=
\sum_{m=1}^8
A_m
\prod_{\substack{k=1\\ k\ne m}}^8
(x-\Lambda_k)
\eeq
is a monic polynomial of degree $7$. The reciprocal required in \cref{eq:recip}, after polynomial division, is
\beq
\frac{1}{\tilde{\xi}_2(z)}
=
\frac{Q(x)}{P(x)} =
x-c_1+\frac{R(x)}{P(x)},
\eeq
where $R(x)$ has degree at most $6$, and
\beq
c_1
=
\sum_{m=1}^8 A_m\Lambda_m.
\eeq
Since $\dot{\Phi}(0)=0$, we have $c_1=0$ as observed in \cref{eq:init-cond-Phi-2}, and therefore
\beq
\frac{1}{\tilde{\xi}_2(z)}
=
x+\frac{R(x)}{P(x)}
=
z-\lambda_2+\frac{R(z-\lambda_2)}{P(z-\lambda_2)}.
\eeq
Substituting this expression into \cref{eq:kernel-s-domain} gives
\beq
\begin{aligned}
\tilde{k}_2(z)
&=
z-\lambda_2
-
\left[
z-\lambda_2
+
\frac{R(z-\lambda_2)}{P(z-\lambda_2)}
\right]
\\
&=
-\frac{R(z-\lambda_2)}{P(z-\lambda_2)}.
\end{aligned}
\eeq

Finally, suppose the roots of $P(x)$ are simple and denote them by $\zeta_k$, $k=1,\ldots,7$. Heaviside's partial-fraction expansion~\cite{cohen_heaviside_1922,doetsch_introduction_1974} gives
\beq
-\frac{R(x)}{P(x)}
=
-\sum_{k=1}^7
\frac{R(\zeta_k)}{P'(\zeta_k)}
\frac{1}{x-\zeta_k}.
\eeq
Since $x=z-\lambda_2$, the inverse Laplace transform is
\beq
k_2(t)
=
-\sum_{k=1}^7
\frac{R(\zeta_k)}{P'(\zeta_k)}
e^{(\lambda_2+\zeta_k)t}.
\eeq
This gives a residue representation of the transverse memory kernel for the fitted three-spectator model. More generally, for $N$ spectators with simple roots, the same procedure reduces the kernel reconstruction to finding the roots of a degree-$(2^N-1)$ polynomial. This avoids a direct numerical Bromwich inversion, although the resulting polynomial root problem can still require numerical conditioning checks.

If some roots of $P(x)$ are repeated rather than simple, the same rational inversion can be performed with the corresponding higher-order partial fractions, producing polynomial prefactors multiplying the exponentials in time.

\subsection{Cancellation of polynomial terms from the initial-time behavior}
\label{app:sec-detailed}

We now show that the leading polynomial terms cancel precisely when the baseline generator $\mathcal L$ captures the instantaneous Markovian contribution to the mode dynamics.
Starting from \cref{eq:mode-memory-equation,eq:xi}, we have
\beq
\dot{\xi}_i(t)
=
\lambda_i\xi_i(t)
+
\int_0^t d\tau\,
k_i(\tau)\xi_i(t-\tau).
\eeq
Let
\beq
I_i(t)
=
\int_0^t d\tau\,
k_i(\tau)\xi_i(t-\tau)
\eeq
denote the memory contribution. If $k_i(\tau)$ is locally integrable near $\tau=0$ and $\xi_i(t)$ is continuous near $t=0$, then
\beq
\begin{aligned}
|I_i(t)|
&\le
\sup_{0\le u\le t}|\xi_i(u)|
\int_0^t d\tau |k_i(\tau)|.
\end{aligned}
\eeq
The supremum is finite by continuity of $\xi_i$, and the integral tends to zero as $t\to0^+$ by local integrability of $k_i$. Hence
\beq
\lim_{t\to0^+}I_i(t)=0.
\eeq
Taking the right limit of the mode equation at the origin gives
\beq
\dot{\xi}_i(0^+)
=
\lambda_i\xi_i(0)+0
=
\lambda_i.
\eeq
Thus, for an ordinary locally integrable memory kernel, the convolution term cannot contribute instantaneously at the initial time: the integration interval has zero length. In this case,
\beq
a_i=\dot{\xi}_i(0^+)=\lambda_i,
\eeq
and the reconstructed $\tilde{k}_i(z)$ has no polynomial part.

The caveat is that this conclusion assumes that all instantaneous Markovian contributions have already been included in the baseline generator $\mathcal L$. If the kernel contains a singular component at the origin,
\beq
k_i(t)=c_i\delta(t)+k_i^{\mathrm{reg}}(t),
\eeq
where $k_i^{\mathrm{reg}}(t)$ is locally integrable, then the convolution contains an instantaneous term proportional to $c_i\xi_i(t)$. The mode equation is then effectively
\beq
\dot{\xi}_i(t)
=
(\lambda_i+c_i)\xi_i(t)
+
\int_0^t d\tau
k_i^{\mathrm{reg}}(\tau)\xi_i(t-\tau),
\eeq
and hence
\beq
\dot{\xi}_i(0^+)=\lambda_i+c_i.
\eeq
In Laplace space, this is exactly the constant polynomial part
\beq
a_i-\lambda_i=c_i
\eeq
that remains in $\tilde{k}_i(z)$. Since this term is instantaneous rather than a finite-memory correction, it is more naturally absorbed into the baseline generator by shifting
\beq
\lambda_i\mapsto\lambda_i+c_i.
\eeq
After this redefinition, the remaining kernel is regular at the origin and represents the genuinely non-Markovian part of the reduced dynamics.

\end{document}